\documentclass[aps,prd,onecolumn,superscriptaddress,preprintnumbers,floatfix,reprint]{revtex4}
\usepackage{graphicx}
\usepackage{amsmath}
\usepackage{bm}
\usepackage{yhmath}
\usepackage[colorlinks,citecolor=blue,urlcolor=blue,linkcolor=blue]{hyperref}
\usepackage{subfigure}
\usepackage{color}
\usepackage{cases}
\usepackage{subfigure}
\usepackage{dcolumn,booktabs,bm}
\usepackage{slashed}
\usepackage{amsfonts,amssymb,stmaryrd,latexsym,amsmath}
\usepackage{textcomp}
 \usepackage{multirow}

\usepackage{array}
\newcounter{RomanNumber}

\newcommand{\lyxmathsym}[1]{\ifmmode\begingroup\def\b@ld{bold}
  \text{\ifx\math@version\b@ld\bfseries\fi#1}\endgroup\else#1\fi}

\def\rmII{{\rm I\!I}}
\def\rmIII{{\rm I\!I\!I}}

\allowdisplaybreaks

\begin{document}

\title{Magnetic moments of the spin-${3\over 2}$ singly heavy baryons}

\author{Lu Meng}\email{lmeng@pku.edu.cn}
\affiliation{School of Physics and State Key Laboratory of Nuclear
Physics and Technology, Peking University, Beijing 100871, China}

\author{Guang-Juan Wang}\email{wgj@pku.edu.cn}\affiliation{School of Physics and State Key Laboratory of Nuclear Physics and Technology, Peking University, Beijing 100871, China}

\author{Chang-Zhi Leng}\email{lengchangzhi@pku.edu.cn}\affiliation{School of Physics and State Key Laboratory of Nuclear Physics and Technology, Peking University, Beijing 100871, China}


\author{Zhan-Wei Liu}\email{liuzhanwei@lzu.edu.cn}
\affiliation{School of Physical Science and Technology, Lanzhou
University, Lanzhou 730000, China}

\author{Shi-Lin Zhu}\email{zhusl@pku.edu.cn}
\affiliation{School of Physics and State Key Laboratory of Nuclear
Physics and Technology, Peking University, Beijing 100871,
China}\affiliation{Collaborative Innovation Center of Quantum
Matter, Beijing 100871, China}

\begin{abstract}
We calculate the magnetic moments of spin-$3\over 2$ singly charmed
baryons in the heavy baryon chiral perturbation theory (HBChPT). The
analytical expressions are given up to $\mathcal{O}(p^3)$. The heavy
quark symmetry is used to reduce the number of low energy constants
(LECs). With the lattice QCD simulation data as the magnetic moments
of the charmed baryons, the numerical results are given up to
$\mathcal{O}(p^3)$ in three scenarios. In the first scenario, we use
the results in the quark model as the leading order input. In the
second scenario, we use the heavy quark symmetry and neglect the
contribution of heavy quark. In the third scenario, the heavy quark
contribution is considered  on the basis of the scenario II and the
magnetic moments of singly bottom baryons are given as a by-product.
\end{abstract}

\maketitle

\thispagestyle{empty}

\section{Introduction}\label{Itro}
The singly heavy baryon contains a heavy  quark and two light
quarks. The two light quarks form the $\bar{3}_f$ and the $6_f$
representation in the SU(3) flavor symmetry.  With the constraint of
Fermi-Dirac statistics, the spin of the $\bar{3}_f$ and the $6_f$
diquarks are $0$ and $1$, respectively. Thus, the total spin of the
$\bar{3}_f$ heavy baryon is $1\over 2$ while that of the $6_f$ heavy
baryon is either $1\over 2$ or $3\over 2$.

The electromagnetic form factors are important properties of the
hadrons, which can reveal their inner structures. The magnetic
moments of hadrons especially attract much attention from the
theorists and
experimentalists~\cite{Jenkins:1994md,Leinweber:1998ej,HackettJones:2000qk,Geng:2008mf,Xiao:2018rvd,Wang:2016dzu,Wang:2017dce}.
The magnetic moments of the singly charmed baryons were investigated
in naive quark model in Ref.~\cite{Lichtenberg:1976fi}. In
Ref.~\cite{JuliaDiaz:2004vh}, the relativistic effect was
considered. The magnetic moments and charge radii of the charmed
baryons are calculated~\cite{JuliaDiaz:2004vh}. The SU(4) chiral
constituent quark model was also adopted to calculated the
(transition) magnetic moments of spin-$1\over 2$ and spin-$3\over 2$
charmed baryons~\cite{Sharma:2010vv}. The masses and magnetic
moments of heavy flavor baryons were calculated in hyper central
model in Ref.~\cite{Patel:2007gx}. The magnetic moments of
spin-$3\over 2$ heavy baryons were obtained using the effective mass
and screened charge scheme~\cite{Dhir:2009ax}. Besides the above
quark models, the MIT bag model was employed to get the magnetic
moments of heavy baryons~\cite{Bose:1980vy}, which were reexamined
in Ref.~\cite{Bernotas:2012nz}. The magnetic moments of charmed
baryons were calculated in the skyrmion
description~\cite{Oh:1991ws}. The mass and magnetic moments of the
heavy flavored baryons were calculated in the QCD sum
rules~\cite{Zhu:1997as,Aliev:2008sk,Ozdem:2018uue}. The magnetic
moments of the lowest-lying singly heavy baryons were investigated
in the chiral quark-soliton model\cite{Yang:2018uoj}. The
(transition) magnetic moments and charge radii of charmed baryons
were simulated with the lattice QCD
recently~\cite{Can:2013tna,Can:2015exa,Bahtiyar:2015sga,Bahtiyar:2016dom}.

The chiral perturbation theory (ChPT) is a model-independent method
to study the hadron
properties~\cite{Weinberg:1978kz,Gasser:1983yg,Gasser:1984gg}.  When
one performs the ChPT in the baryon sector, the nonvanashing baryon
mass in chiral limit will mess up the power counting used in the
pure meson sector. The heavy baryon chiral perturbation theory
(HBChPT) was introduced to solve the
problem~\cite{Jenkins:1990jv,Bernard:1992qa}. The HBChPT is expanded
by the momenta of pseudoscalar mesons and the residual momenta of
heavy baryons. The HBChPT was widely performed to calculated the
electromagnetic properties of baryons. The magnetic moments of octet
and decuplet baryons were calculated in HBChPT
scheme~\cite{Jenkins:1992pi,Bernard:1995dp,Meissner:1997hn,Kubis:2000aa,Puglia:1999th,Li:2016ezv}.
The (transition) magnetic moments of doubly heavy baryons were
investigated in Refs.~\cite{Li:2017cfz,Li:2017pxa,Meng:2017dni}. The
magnetic moments of singly charmed baryons were calculated up to the
next-to-next-to-leading order in
HBChPT\cite{Banuls:1999mu,Zamiralov:2001ix}. In our recent work, we
calculated the magnetic moments of spin-$1\over 2$ singly charmed
baryons up to the $\mathcal{O}(p^3)$~\cite{Wang:2018gpl}.

The dynamics of singly heavy hadron is constrained by both the
chiral symmetry in light quark sector and heavy quark symmetry in
heavy quark sector. The heavy quark symmetry and the chiral symmetry
were often combined to investigate the singly heavy hadrons. In
Ref.~\cite{Yan:1992gz}, the authors constructed the chiral
Lagrangians of heavy mesons ($Q\bar{q}$) and heavy baryons ($Qqq$)
and calculated their strong and semileptonic weak decays
incorporating with heavy quark symmetry. The decay properties of
singly heavy hadrons were calculated in a formalism which combines
the chiral symmetry and the heavy quark
symmetry~\cite{Cheng:1992xi,Cheng:1993gc,Cheng:1993kp,Wise:1992hn,Burdman:1992gh,Cho:1992nt}.
The electromagnetic decays of $D_{s0}(2317)$ and $D_{s1}(2460)$ are
investigated in the heavy-hadron chiral perturbation theory with the
heavy quark symmetry~\cite{Mehen:2004uj}.

 In this work, we calculate the magnetic moments of the spin-$3\over 2$ singly heavy baryons in the HBChPT scheme. In Section~\ref{SecEM}, we perform the multiple expansion of the electromagnetic current matrix element for spin-$3\over 2$ baryons. In Section~\ref{SecLag}, we construct the Lagrangians used in calculating the magnetic moments. In Section~\ref{Secanal}, we calculate the analytical expressions of the magnetic moments order by order up to $\mathcal{O}(p^3)$. In Section~\ref{HQSS}, we reduce the numbers of independent LECs in our analytical results with the heavy quark spin symmetry. We give the numerical results in three scenarios in Section~\ref{nrslt}. Some discussions and a brief conclusion are given in the Section~\ref{Disscss}. The integrals used in this work and some
  by-products are listed in the Appendix.

\section{Electromagnetic form factors of the spin-$\frac{3}{2}$ baryons }\label{SecEM}
Constrained by the time reversal (T), the parity (P), charge
conjugate (C) and the gauge invariance, the matrix element of the
electromagnetic current for spin-$3\over 2$ particles takes the
following form~\cite{Nozawa:1990gt,Ledwig:2011cx},
\begin{equation}
\langle
T(p^{\prime})|J_{\mu}|T(p)\rangle=\bar{u}^{\rho}(p^{\prime})O_{\rho\mu\sigma}(p^{\prime},p)u^{\sigma}(p),
\end{equation}
with
\begin{equation}
O_{\rho\mu\sigma}(p^{\prime},p)=-g_{\rho\sigma}\left[\gamma_{\mu}F_{1}(q^{2})+\frac{i\sigma_{\mu\alpha}q^{\alpha}}{2M_{T}}F_{2}(q^{2})\right]-\frac{q_{\rho}q_{\sigma}}{4M_{T}{}^{2}}\left[\gamma_{\mu}F_{3}(q^{2})+\frac{i\sigma_{\mu\alpha}q^{\alpha}}{2M_T}F_{4}(q^{2})\right],
\end{equation}
where $p$ and $p'$ are the momenta of the spin-$3\over 2$ baryons.
$P=p+p'$, $q=p'-p$. $M_T$ is the baryon mass and $u_{\sigma}$ is the
Rarita-Schwinger spinor~\cite{Rarita:1941mf}.

The charge (E0), electro-quadrupole (E2), magnetic-dipole (M1), and
magnetic octupole (M3) form factors read
\begin{equation}
\begin{cases}
{ G_{E0}(q^{2})=F_{1}-\tau F_{2}+\frac{2}{3}\tau G_{E2},}\\
{ G_{E2}(q^{2})=F_{1}-\tau F_{2}-\frac{1}{2}(1+\tau)(F_{3}-\tau F_{4}),}\\
{ G_{M1}(q^{2})=F_{1}+F_{2}+\frac{4}{5}\tau G_{M3},}\\
{ G_{M3}(q^{2})=F_{1}+F_{2}-\frac{1}{2}(1+\tau)(F_{3}+F_{4}).}
\end{cases}
\end{equation}
where $\tau=-\frac{q^{2}}{(2M_{T})^{2}}$. On the right-hand side, we
omit the variable $q^2$ of $F_i$ for convenience. The
magnetic-dipole form factor is related to the magnetic moment as
\begin{eqnarray}
\mu_{T}=G_{M1}(0){e\over 2M_T}.
\end{eqnarray}

In HBChPT scheme, The baryon momentum $p^{\mu}$ is decomposed into
the $M_Tv^{\mu}$ and a residual momentum $k^{\mu}$, where $v_{\mu}$
is the velocity of the baryon and $v^2=1$. The baryon field $T$ is
decomposed into a ``light" field $\mathcal{T}(x)$ and a ``heavy"
field $\mathcal{N}(x)$,
\begin{eqnarray}
&\mathcal{T}(x)=e^{iM_Tv\cdot x}{1+\slashed{v}\over 2}T(x),\\
&\mathcal{N}(x)=e^{iM_Tv\cdot x}{1-\slashed{v}\over 2}T(x).
\end{eqnarray}
After integrating out the heavy degrees of freedom, one gets the
nonrelativistic Lagrangians. In the HBChPT scheme, the theory is
expanded by either the momenta of the pseudoscalar mesons or the
residual momenta of the baryons.

In the HBChPT scheme, the matrix element of the electromagnetic
current $J_{\mu}$ is reduced as \cite{Li:2016ezv}
 \begin{equation}
\langle
\mathcal{T}(p^{\prime})|J_{\mu}|\mathcal{T}(p)\rangle=\bar{u}^{\rho}(p^{\prime})\mathcal{O}_{\rho\mu\sigma}(p^{\prime},p)u^{\sigma}(p),
\end{equation}
with
 \begin{equation}
 \mathcal{O}_{\rho\mu\sigma}(p^{\prime},p)=-g_{\rho\sigma}\left[v_{\mu}\left(F_{1}-\tau F_{2}\right)+\frac{[S_{\mu},S_{\alpha}]}{M_{T}}q^{\alpha}\left(F_{1}+F_{2}\right)\right]-\frac{q^{\rho}q^{\sigma}}{4M_{T}^{2}}\left[v_{\mu}\left(F_{3}-\tau F_{4}\right)+\frac{[S_{\mu},S_{\alpha}]}{M_{T}}q^{\alpha}\left(F_{3}+F_{4}\right)\right],
  \end{equation}
where $S_{\mu}={i\over 2}\gamma_5\sigma_{\mu\nu}v^\nu$ is the
covariant spin-operator.

\begin{figure}
\centering
\includegraphics[scale=1.0]{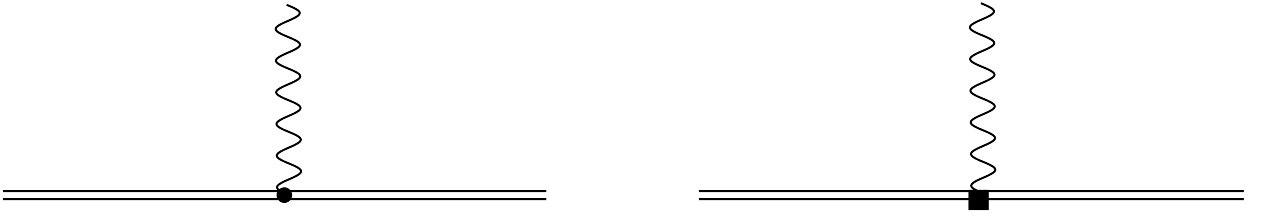}
\caption{The tree diagrams contribute to the magnetic moments of the
spin-$3\over 2$ heavy baryon. The solid dot and black square
represents $\mathcal{O}(p^2)$ and $\mathcal{O}(p^4)$ vertices,
representatively. }\label{tree}
\end{figure}

The tree and loop Feynman diagrams contributing to the magnetic
moments are shown in Figs.~\ref{tree} and~\ref{loop}, respectively.
According to the standard power
counting~\cite{Scherer:2002tk,Bernard:1995dp}, the chiral order
$D_{\chi}$ of a Feynmen digram is
\begin{equation}
D_{\chi}=2L+1+\sum_d (d-2)N_d^{\phi}+\sum_d(d-1)N_d^{\phi B}
\end{equation}
where $L$, $N_d^{\phi}$ and $N_d^{\phi B}$ are the numbers of loops,
pure meson vertices and meson-baryon vertices, respectively. $d$ is
the chiral dimension. The chiral order of the magnetic moment
$\mu_T$ is counted as $(D_{\chi}-1)$.

\section{Chiral Lagrangians}\label{SecLag}

\subsection{The leading order chiral Lagrangians}
We choose the nonlinear realization of the chiral symmetry,
\begin{eqnarray}
U=u^{2}=\exp(i\phi/F_{0}),
\end{eqnarray}
where $\phi$ is the matrix for octet Goldstones,
\begin{equation}
\phi=\left(\begin{array}{ccc}
\pi^{0}+\frac{1}{\sqrt{3}}\eta & \sqrt{2}\pi^{+} & \sqrt{2}K^{+}\\
\sqrt{2}\pi^{-} & -\pi^{0}+\frac{1}{\sqrt{3}}\eta & \sqrt{2}K^{0}\\
\sqrt{2}K^{-} & \sqrt{2}\bar{K}^{0} & -\frac{2}{\sqrt{3}}\eta
\end{array}\right),
\end{equation}
 $F_0$ is the decay constant of the pseudoscalar meson in chiral limit. We adopt $F_{\pi}=$92.4 MeV, $F_{K}=$ 113 MeV and $F_{\eta}=$ 116 MeV in this work.
Under the $\mbox{SU(3)}_L\times \mbox{SU(3)}_R$ chiral
transformation, the $U$ and $u$ respond according to
\begin{eqnarray}
U&\rightarrow & R U L^\dagger,\\
u& \rightarrow & RuK^\dagger=KuL^\dagger,
\end{eqnarray}
where $R$ and $L$ are $\mbox{SU(3)}_R$ and $\mbox{SU(3)}_L$
transformation matrices, respectively. $K=K(R,L,\phi)$ is a unitary
transformation.

We use the notations $B_{\bar{3}}$, $B_6$ and $B_6^*$ to denote the
antitriplet, spin-$1\over 2$ sextet and spin-$3\over 2 $ sextet,
respectively. These baryon fields are realized as~\cite{Yan:1992gz}:
\begin{equation}
B_{\bar{3}}=\left(\begin{array}{ccc}
0 & \Lambda_{c}^{+} & \Xi_{c}^{+}\\
-\Lambda_{c}^{+} & 0 & \Xi_{c}^{0}\\
-\Xi_{c}^{+} & -\Xi_{c}^{0} & 0
\end{array}\right),\quad B_{6}=\left(\begin{array}{ccc}
\Sigma_{c}^{++} & \frac{\Sigma_{c}^{+}}{\sqrt{2}} & \frac{\Xi_{c}^{\prime+}}{\sqrt{2}}\\
\frac{\Sigma_{c}^{+}}{\sqrt{2}} & \Sigma_{c}^{0} & \frac{\Xi_{c}^{\prime0}}{\sqrt{2}}\\
\frac{\Xi_{c}^{\prime+}}{\sqrt{2}} &
\frac{\Xi_{c}^{\prime0}}{\sqrt{2}} & \Omega_{c}^{0}
\end{array}\right),\quad B_{6}^{*\mu}=\left(\begin{array}{ccc}
\Sigma_{c}^{*++} & \frac{\Sigma_{c}^{*+}}{\sqrt{2}} & \frac{\Xi_{c}^{*+}}{\sqrt{2}}\\
\frac{\Sigma_{c}^{*+}}{\sqrt{2}} & \Sigma_{c}^{*0} & \frac{\Xi_{c}^{*0}}{\sqrt{2}}\\
\frac{\Xi_{c}^{*+}}{\sqrt{2}} & \frac{\Xi_{c}^{*0}}{\sqrt{2}} &
\Omega_{c}^{*0}
\end{array}\right)^{\mu}
\end{equation}
The chiral transformation can be established:
\begin{equation}
B\rightarrow KBK^T \label{chiraltrans}
\end{equation}
where $B$ represents the $B_{\bar{3}}$, the $B_6$ or the $B_6^*$
field.

We introduce the left-handed and the right-handed external fields as
the electromagnetic fields:
\begin{equation}
r_{\mu}=l_{\mu}=-eQ_{m(c)}A_{\mu},
\end{equation}
where $A_{\mu}$ is the electromagnetic field and $Q_{m(c)}$
represents the meson (singly charmed baryon) charge matrix. In this
work, $Q_m=\text{diag}(2/3,-1/3,-1/3)$ and $Q_c=\text{diag}(1,0,0)$.


We define some ``building blocks" before constructing Lagrangians.
The chiral connection and axial vector field are defined
as~\cite{Scherer:2002tk,Bernard:1995dp},
\begin{equation}
\Gamma_{\mu}=\frac{1}{2}\left[u^{\dagger}(\partial_{\mu}-ir_{\mu})u+u(\partial_{\mu}-il_{\mu})u^{\dagger}\right],
\end{equation}
\begin{equation}
u_{\mu}= {i\over
2}\left[u^{\dagger}(\partial_{\mu}-ir_{\mu})u-u(\partial_{\mu}-il_{\mu})u^{\dagger}\right],
\end{equation}
The chiral covariant QED field strength tensors $F_{\mu\nu}^{\pm}$
are defined as
\begin{eqnarray}
F_{\mu\nu}^{\pm} & = & u^{\dagger}F_{\mu\nu}^{R}u\pm
uF_{\mu\nu}^{L}u^{\dagger},\\
F_{\mu\nu}^{R} & = &
\partial_{\mu}r_{\nu}-\partial_{\nu}r_{\mu}-i[r_{\mu},r_{\nu}],\\
F_{\mu\nu}^{L} & = &
\partial_{\mu}l_{\nu}-\partial_{\nu}l_{\mu}-i[l_{\mu},l_{\nu}].
\end{eqnarray}
In order to introduce the chiral symmetry breaking effect, we define
$\chi_{\pm}$,
\begin{eqnarray}
&\chi_{\pm}=u^{\dagger}\chi u^\dagger \pm u\chi^\dagger u,\nonumber \\
&\chi=2B_0~\text{diag}(m_u,m_d,m_s)
\end{eqnarray}
where $B_0$ is a parameter related to the quark condensate and
$m_{u,d,s}$ is the current quark mass.

The leading order ($\mathcal{O}(p^{2})$) pure-meson Lagrangian is
\begin{equation}
\mathcal{L}_{\phi\phi}^{(2)}=\frac{F_{0}^{2}}{4}\langle\nabla_{\mu}U(\nabla^{\mu}U)^{\dagger}\rangle
\label{1phigamma},
\end{equation}
 where the superscript denotes the chiral order. The $\langle X\rangle$ means the trace of field $X$. The covariant derivative of Goldstone fields is define as
\begin{equation}
\nabla_{\mu}U=\partial_{\mu}U-ir_{\mu}U+iUl_{\mu}.
\end{equation}
The leading order Lagrangians for singly heavy baryons read
\begin{eqnarray}
\mathcal{L}_{B\phi}^{(1)}&=&\frac{1}{2}\langle\bar{B}_{\bar{3}}\left(i\slashed{D}-M_{\bar{3}}\right)B_{\bar{3}}\rangle+\langle\bar{B}_{6}(i\slashed{D}-M_{6})B_{6}\rangle \nonumber\\
&+& \langle\bar{B}_{6}^{*\mu}[-g_{\mu\nu}(i\slashed{ D}-M_{6^*})+i(\gamma_{\mu}D_{\nu}+\gamma_{\nu}D_{\mu})-\gamma_{\mu}(i\slashed{D}+M_{6^*})\gamma_{\nu}]B_{6}^{*\nu}\rangle \nonumber\\
&+&g_{1}\langle\bar{B}_{6}\gamma_{\mu}\gamma_{5}u^{\mu}B_{6}\rangle+g_{2}\langle\bar{B}_{6}\gamma_{\mu}\gamma_{5}u^{\mu}B_{\bar{3}}+{\rm H.c.}\rangle+g_{3}\langle\bar{B}_{6\mu}^{*}u^{\mu}B_{6}+{\rm H.c.}\rangle \nonumber \\
&+&g_{4}\langle\bar{B}_{6\mu}^{*}u^{\mu}B_{\bar{3}}+{\rm
H.c.}\rangle+g_{5}\langle\bar{B}_{6}^{*\nu}\gamma_{\mu}\gamma_{5}u^{\mu}B_{6\nu}^{*}\rangle+g_{6}\langle\bar{B}_{\bar{3}}\gamma_{\mu}\gamma_{5}u^{\mu}B_{\bar{3}}\rangle,
\end{eqnarray}
where $g_i$ is the axial charge. In this work, we ignore the mass
splitting among the particles in the same multiplet. $M_{\bar{3}}$,
$M_6$ and $M_{6^*}$ are the average baryon masses for  the
antritriplet, spin-$1\over 2$ sextet and spin-$3\over 2$ sextet,
respectively.

In the framework of HBChPT, the leading order nonrelativistic
Lagrangians read
\begin{eqnarray}
\mathcal{L}_{\mathcal{B}\phi}^{(1)}&=&\frac{1}{2}\langle\bar{\mathcal{B}}_{\bar{3}}iv\cdot D\mathcal{B}_{\bar{3}}\rangle+\langle\bar{\mathcal{B}}_{6}(iv\cdot D-\delta_{2})\mathcal{B}_{6}\rangle-\langle\bar{\mathcal{B}}_{6}^{*}(iv\cdot D-\delta_{3})\mathcal{B}_{6}^{*}\rangle \nonumber\\
    &+&2g_{1}\langle\bar{\mathcal{B}}_{6}S\cdot u\mathcal{B}_{6}\rangle+2g_{2}\langle\bar{\mathcal{B}}_{6}S\cdot u\mathcal{B}_{\bar{3}}+{\rm H.c.}\rangle+g_{3}\langle\bar{\mathcal{B}}_{6\mu}^{*}u^{\mu}\mathcal{B}_{6}+{\rm H.c.}\rangle \nonumber \\
    &+&g_{4}\langle\bar{\mathcal{B}}_{6\mu}^{*}u^{\mu}\mathcal{B}_{\bar{3}}+{\rm H.c.}\rangle+2g_{5}\langle\bar{\mathcal{B}}_{6}^{*}S\cdot u\mathcal{B}_{6}^{*}\rangle+2g_{6}\langle\bar{\mathcal{B}}_{\bar{3}}S\cdot u\mathcal{B}_{\bar{3}}\rangle,\label{1Bphi}
\end{eqnarray}
where we ignore the terms suppressed by $1\over M_T$.
$\delta_{1,2,3}$ are the mass differences between different
multiplets,
\begin{eqnarray}
\delta_{1}  =M_{6*}-M_{6},\quad\delta_{2}   =M_{6}-M_{\bar{3}},\quad
\delta_{3}  =M_{6*}-M_{\bar{3}}.
\end{eqnarray}

\subsection{The next-to-leading order chiral Lagrangians}
The $O(p^2)$ baryon-photon Lagrangians contributing to the magnetic
moments read:
\begin{eqnarray}
{\cal L}_{B\gamma}^{(2)}&=&\frac{d_{2}}{8M_{N}}\langle\bar{B}_{3}\sigma^{\mu\nu}\hat{F}_{\mu\nu}^{+}B_{3}\rangle+\frac{d_{3}}{8M_{N}}\langle\bar{B}_{3}\sigma^{\mu\nu}B_{3}\rangle\langle F_{\mu\nu}^{+}\rangle+\frac{d_{5}}{8M_{N}}\langle\bar{B}_{6}\sigma^{\mu\nu}\hat{F}_{\mu\nu}^{+}B_{6}\rangle+\frac{d_{6}}{8M_{N}}\langle\bar{B}_{6}\sigma^{\mu\nu}B_{6}\rangle\langle F_{\mu\nu}^{+}\rangle \nonumber \\
&+&\frac{f_{2}}{8M_{N}}\langle\bar{B}_{3}\sigma^{\mu\nu}\hat{F}_{\mu\nu}^{+}B_{6}\rangle+{\rm H.c.}+\frac{if_{4}}{8M_{N}}\langle\bar{B}_{3}\hat{F}_{\mu\nu}^{+}\gamma^{\nu}\gamma^{5}B_{6}^{*\mu}\rangle+{\rm H.c.}+\frac{if_{6}}{8M_{N}}\langle\bar{B}_{6}\hat{F}_{\mu\nu}^{+}\gamma^{\nu}\gamma^{5}B_{6}^{*\mu}\rangle+{\rm H.c.}  \nonumber\\
&+&\frac{if_{7}}{8M_{N}}\langle\bar{B}_{6}\gamma^{\nu}\gamma^{5}B_{6}^{*\mu}\rangle\langle
F_{\mu\nu}^{+}\rangle+{\rm
H.c.}+\frac{if_{9}}{4M_{N}}\langle\bar{B}_{6}^{*\mu}\hat{F}_{\mu\nu}^{+}B_{6}^{*\nu}\rangle+\frac{if_{10}}{4M_{N}}\langle\bar{B}_{6}^{*\mu}B_{6}^{*\nu}\rangle\langle
F_{\mu\nu}^{+}\rangle, \label{2Bgammac}
\end{eqnarray}
 where $d_i$ and $f_i$ are the coupling constants. $\hat{X}=X-{1\over 3 }\langle X \rangle$ is the traceless part of the field $X$. The $f_9$ and $f_{10}$ terms contribute to the leading order magnetic moments of spin-$3\over 2$ heavy baryons in the tree diagrams. Other terms will contribute to the higher order magnetic moments  in the loop diagrams.
 The nonrelativistic form of Eq.~(\ref{2Bgammac}) reads

\begin{eqnarray}
{\cal L}_{\mathcal{B}\gamma}^{(2)}&=&-\frac{id_{2}}{4M_{N}}\langle\bar{\mathcal{B}}_{3}[S^{\mu},S^{\nu}]\hat{F}_{\mu\nu}^{+}\mathcal{B}_{3}\rangle-\frac{id_{3}}{4M_{N}}\langle\bar{\mathcal{B}}_{3}[S^{\mu},S^{\nu}]\mathcal{B}_{3}\rangle\langle F_{\mu\nu}^{+}\rangle-\frac{id_{5}}{4M_{N}}\langle\bar{\mathcal{B}}_{6}[S^{\mu},S^{\nu}]\hat{F}_{\mu\nu}^{+}\mathcal{B}_{6} \nonumber \\
&-&\frac{id_{6}}{4M_{N}}\langle\bar{\mathcal{B}}_{6}[S^{\mu},S^{\nu}]\mathcal{B}_{6}\rangle\langle F_{\mu\nu}^{+}\rangle \nonumber -\frac{if_{2}}{4M_{N}}\langle\bar{\mathcal{B}}_{3}[S^{\mu},S^{\nu}]\hat{F}_{\mu\nu}^{+}\mathcal{B}_{6}\rangle+{\rm H.c.}+\frac{if_{4}}{4M_{N}}\langle\bar{\mathcal{B}}_{3}F_{\mu\nu}^{+}S^{\nu}\mathcal{B}_{6}^{*\mu}\rangle+{\rm H.c.}\nonumber\\
&+&\frac{if_{6}}{4M_{N}}\langle\bar{\mathcal{B}}_{6}\hat{F}_{\mu\nu}^{+}S^{\nu}\mathcal{B}_{6}^{*\mu}\rangle+{\rm H.c.}+\frac{if_{7}}{4M_{N}}\langle\bar{\mathcal{B}}_{6}S^{\nu}\mathcal{B}_{6}^{*\mu}\rangle\langle F_{\mu\nu}^{+}\rangle+{\rm H.c.}+\frac{if_{9}}{4M_{N}}\langle\bar{\mathcal{B}}_{6}^{*\mu}\hat{F}_{\mu\nu}^{+}\mathcal{B}_{6}^{*\nu}\rangle \nonumber\\
&+&\frac{if_{10}}{4M_{N}}\langle\bar{\mathcal{B}}_{6}^{*\mu}\mathcal{B}_{6}^{*\nu}\rangle\langle
F_{\mu\nu}^{+}\rangle,\label{2Bgamma}
\end{eqnarray}
The $d_2$ terms in Eqs.~(\ref{2Bgammac}) and
(\ref{2Bgamma}) represent the contribution of the light degrees of
freedom to the magnetic moments of the anti-triplet baryons . Since
the $J^P$ of the light diquark in the anti-triplet baryons is $0^+$,
the M1 radiative transition
$|J=0\rangle\rightarrow|J=0\rangle+\gamma$ is forbidden. The light
diquark does not contribute to the magnetic moment and $d_2=0$. 

We also construct the $\mathcal{O}(p^2)$ meson-meson-baryon
interaction Lagrangian as followings, which contributes to the
$O(p^3)$ magnetic moments through the loop (j) in Fig~\ref{loop}.
\begin{eqnarray}
{\cal
L}_{B\phi\phi}^{(2)}=-\frac{f_{8}}{2M_{N}}\langle\bar{B}_6^{*\mu}[u_{\mu},u_{\nu}]B_6^{*\nu}\rangle.\label{2Bphiphi}
\end{eqnarray}

\subsection{The higher order chiral Lagrangians }
 According to the group representation theory, there are seven interaction terms in $\mathcal{O}(p^4)$ Lagrangians which contribute to the $\mathcal{O}(p^3)$ magnetic moments in the tree diagrams~\cite{Wang:2018gpl}. The $\chi_{+}=4B_0~\text{diag}(0,0,m_s)=4B_0m_s~\tilde{\chi}_+$ at the leading order. We use the $\tilde{\chi}_+$ as the building block and the $B_0$ and $m_s$ are absorbed into the LECs. There are only two independent nonvanishing terms \begin{eqnarray}
{\cal
L}_{B\phi}^{(4)}&=&\frac{ih_{2}}{4M_{N}}\langle\bar{B}_{6}^{*\mu}\langle
F_{\mu\nu}^{+}\rangle{\tilde{\chi}}_{+}B_{6}^{*\nu}+\frac{ih_{4}}{M_{N}}\langle\bar{B}_{6}^{*\mu}\hat{F}_{\mu\nu}^{+}B_{6}^{*\nu}{\tilde{\chi}}^{T}_+\rangle.\label{4Bgamma}
\end{eqnarray}


\section{Analytical expressions}\label{Secanal}

\begin{figure}[htb]
  \centering
  \includegraphics[scale=1.0]{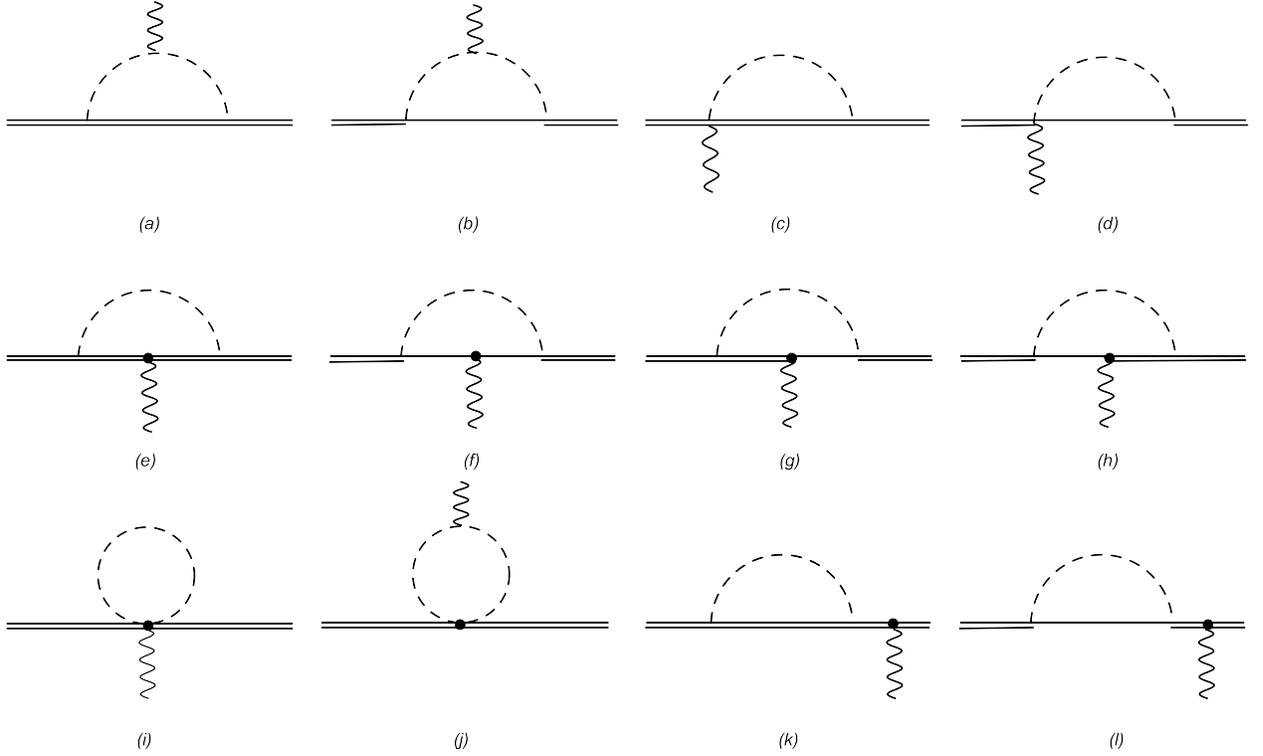}
  \caption{The loop diagrams contribute to the magnetic moments of the spin-$3\over 2$ heavy baryons. The single and double lines represent the spin-$1\over 2$ and spin-$3\over 2$ heavy baryons, respectively. The solid dots denote the next-leading order vertices, while the other vertices are at the leading order. The diagrams (a)-(d) contribute to the $\mathcal{O}(p^2)$ magnetic moments, while the (e)-(l) diagrams contribute to the $\mathcal{O}(p^3)$ magnetic moments.}\label{loop}
  \end{figure}

The leading order magnetic moments are at $\mathcal{O}(p)$, which
stem from $\mathcal{O}(p^2)$ vertices in Eq. (\ref{2Bgamma}):
\begin{equation}
\mu_{\Sigma_{c}^{*++}}^{(1)}=-\left({2\over3}f_{9}+f_{10}\right)\mu_{N},\quad\mu_{\Sigma_{c}^{*+}}^{(1)}=\mu_{\Xi_{c}^{*+}}^{(1)}=-\left(\frac{1}{6}f_{9}+f_{10}\right)\mu_{N},\quad\mu_{\Sigma_{c}^{*0}}^{(1)}=\mu_{\Xi_{c}^{*0}}^{(1)}=\mu_{\Omega_{c}^{*0}}^{(1)}=\left({1\over3}f_9-f_{10}\right)\mu_{N}.\label{mu1}
\end{equation}
There are two unknown LECs $f_9$ and $f_{10}$ at this order.

Four loop diagrams (a)-(d) in Fig.~\ref{loop} contribute to the
$\mathcal{O}(p^2)$ magnetic moments. The meson-photon vertex arises
from the $\mathcal{L}_{\phi\phi}^{(2)}$, while the meson-baryon
vertex is from the $\mathcal{L}_{\mathcal{B}\phi}^{(1)}$.  The
diagrams (c) and (d) vanish for the structure $v_{\mu} u^{\mu}$ in
the amplitude~\cite{Li:2016ezv,Meng:2017dni}. The corrections from
the loops (a)-(d) read
\begin{eqnarray}
&&\mu^{(2,a)}=\beta^{\phi}\frac{g_{5}^{2}M_{N}}{2F_{\phi}^{2}}\frac{3-d}{d-1}n_{1}^{II}(0,m_{\phi})\mu_{N},\\
&&\mu^{(2,b)}=-\beta^{\phi}\frac{g_{3}^{2}M_{N}}{4F_{\phi}^{2}}n_{1}^{II}(\delta_{1},m_{\phi})\mu_{N}-2\beta^{\phi}\frac{g_{4}^{2}M_{N}}{4F_{\phi}^{2}}n_{1}^{II}(\delta_{3},m_{\phi})\mu_{N},\label{mu2}
\end{eqnarray}
where the $n^{\rmII}_{1}(\omega,m_\phi)$ is the loop integral given
in Appendix~\ref{integrals}. The $\beta^{\phi}$ is the coefficient
in Table~\ref{cg}. There exist three LECs $g_{3,4,5}$ to be
determined at this order.

The $\mathcal{O}(p^3)$ magnetic moments come from both the tree
diagrams and the loop diagrams. The vertices of tree diagrams are
from the interaction in Eq.~(\ref{4Bgamma}). The results  of the
tree diagram read,
\begin{eqnarray}
&\mu_{\Sigma_{c}^{*++}}^{(3,\text{tree})}=
\mu_{\Sigma_{c}^{*+}}^{(3,\text{tree})}=
\mu_{\Sigma_{c}^{*0}}^{(3,\text{tree})}=0,\nonumber \\
&\mu_{\Xi_{c}^{*+}}^{(3,\text{tree})}=-\left(\frac{1}{2}h_{2}+{4\over
3}h_{4}\right)\mu_{N},
\quad\mu_{\Xi_{c}^{*0}}^{(3,\text{tree})}=-\left(\frac{1}{2}h_{2}-{2\over
3}h_{4}\right)\mu_{N},
\quad\mu_{\Omega_{c}^{*0}}^{(3,\text{tree})}=-\left(h_{2}-{4\over
3}h_{4}\right)\mu_{N}.\label{mu3tree}
\end{eqnarray}

The loop diagrams (e)-(l) in Fig.~\ref{loop} contribute to the
$\mathcal{O}(p^3)$ magnetic moments. The baryon-photon vertices in
loop diagrams (e)-(h) come from the
$\mathcal{L}^{(2)}_{\mathcal{B}\gamma}$ in Eq.~(\ref{2Bgamma}). The
baryon-meson vertices are from the axial coupling Lagrangian in
Eq.~(\ref{1Bphi}). The vertex in the loop diagram comes from the
$f_9$ term in Eq.~(\ref{2Bgamma}). The meson-meson-baryon vertex in
the loop diagram (j) comes from the interaction (\ref{2Bphiphi}).
Diagrams (k) and (l) are the renormalization of the spin-$3\over 2$
baryon fields. The $\mathcal{O}(p^3)$ corrections from the above
loop diagrams read,
\begin{eqnarray}
&&\mu^{(3,i)}=2\beta^{\phi}\frac{f_{9}m_{\phi}^{2}}{128\pi^{2}F_{\phi}^{2}}\text{ln}\frac{m_{\phi}^{2}}{\lambda^{2}}\mu_{N},\\
&&\mu^{(3,j)}=-4\beta^{\phi}\frac{f_{8}m^{2}_{\phi}}{256\pi^{2}F_{\phi}^{2}}\text{ln}\frac{m^{2}_{\phi}}{\lambda^{2}}\mu_{N},\\
&&\mu^{(3,e)}=\gamma_{1}^{\phi}\frac{g_{5}^{2}}{2F_{\phi}^{2}}\left(\frac{1-d}{2}+\frac{4}{d-1}-\frac{4}{\left(d-1\right)^{2}}\right)J_{2}'(0)\mu_{N},\\
&&\mu^{(3,f)}=\gamma_{2}^{\phi}\frac{g_{3}^{2}}{4F_{\phi}^{2}}J_{2}'(\delta_{1})\text{\ensuremath{\mu}}_{N}+\rho^{\phi}\frac{g_{4}^{2}}{4F_{\phi}^{2}}J_{2}'(\delta_{3})\mu_{N}+2\delta^{\phi}\frac{f_{2}g_{3}g_{4}}{4F_{\phi}^{2}}\frac{J_{2}(\delta_{1})-J_{2}(\delta_{3})}{\delta_{1}-\delta_{3}}\mu_{N},\\
&&\mu^{(3,g)}=\mu^{(3,h)}=\gamma_{3}^{\phi}\frac{g_{5}g_{3}}{4F_{\phi}^{2}}\left(\frac{d-3}{d-1}\right)\frac{J_{2}(\delta_{1})-J_{2}(0)}{(-\delta_{1})}\mu_{N}+\delta^{\phi}\frac{g_{5}g_{4}f_{4}}{4F_{\phi}^{2}}\left(\frac{d-3}{d-1}\right)\frac{J_{2}(\delta_{3})-J_{2}(0)}{(-\delta_{3})}\mu_{N},\\
&&\mu^{(3,k)}=\xi^{\phi}\frac{g_{5}^{2}}{F_{\phi}^{2}}J_{2}'(0)\left(\frac{1-d}{4}+\frac{1}{d-1}\right)\mu^{(1)},\\
&&\mu^{(3,l)}=\xi^{\phi}\frac{-g_{3}^{2}}{4F_{\phi}^{2}}J_{2}'(\delta_{1})\mu^{(1)}+\eta^{\phi}\frac{-g_{4}^{2}}{4F_{\phi}^{2}}J_{2}'(\delta_{3})\mu^{(1)},
\label{mu3loop}
\end{eqnarray}
where $\beta^{\phi}$, $\gamma_i^{\phi}$, $\rho^{\phi}$,
$\delta^{\phi}$, $\xi^{\phi}$ and $\eta^{\phi}$ are the coefficients
of loops, which are given in Table~\ref{cg}. There are thirteen new
LECs introduced at this order.

\begin{table}
  \caption{The coefficients of the loop diagrams in Fig~\ref{loop}.}\label{cg}
\begin{tabular}{l|c|cccccc}
\toprule[1pt]\toprule[1pt] Loop &  & $\Sigma_{c}^{*++}$ &
$\Sigma_{c}^{*+}$ & $\Sigma_{c}^{*0}$ & $\Xi_{c}^{*+}$ &
$\Xi_{c}^{*0}$ & $\Omega_{c}^{*0}$\tabularnewline \midrule[1pt]
(a),(b), & $\beta^{\pi}$ & $2$ &  & $-2$ & $1$ & $-1$ &
\tabularnewline (i),(j) & $\text{\ensuremath{\beta}}^{K}$ & $2$ &
$1$ &  &  & $-1$ & $-2$\tabularnewline\midrule[1pt]
\multirow{11}{*}{(e)-(h)} & $\gamma_{1}^{\pi}$ &
$\frac{5}{6}f_{9}+2f_{10}$ & $\frac{1}{3}f_{9}+2f_{10}$ &
$-\frac{1}{6}f_{9}+2f_{10}$ & $-\frac{1}{8}f_{9}+\frac{3}{4}f_{10}$
& $\frac{3}{4}f_{10}$ & \tabularnewline
 & $\gamma_{1}^{K}$ & $\frac{1}{6}f_{9}+f_{10}$ & $-\frac{1}{12}f_{9}+f_{10}$ & $-\frac{1}{3}f_{9}+f_{10}$ & $\frac{5}{12}f_{9}+\frac{5}{2}f_{10}$ & -$\frac{7}{12}f_{9}+\frac{5}{2}f_{10}$ & $-\frac{1}{6}f_{9}+2f_{10}$\tabularnewline
  & $\gamma_{1}^{\eta}$ & $\frac{2}{9}f_{9}+\frac{1}{3}f_{10}$ & $\frac{1}{18}f_{9}+\frac{1}{3}f_{10}$ & $-\frac{1}{9}f_{9}+\frac{1}{3}f_{10}$ & $\frac{1}{72}f_{9}+\frac{1}{12}f_{10}$ & $-\frac{1}{36}f_{9}+\frac{1}{12}f_{10}$ & $-\frac{4}{9}f_{9}+\frac{4}{3}f_{10}$\tabularnewline
\cline{2-8}
 & $\gamma_{2}^{\phi}$ & \multicolumn{6}{c}{$\gamma_{2}^{\phi}=\gamma_{1}^{\phi}$ $(f_{9}\rightarrow d_{5},f_{10}\rightarrow d_{6})$}\tabularnewline
\cline{2-8}
 & $\gamma_{3}^{\phi}$ & \multicolumn{6}{c}{$\gamma_{3}^{\phi}=\gamma_{1}^{\phi}(f_{9}\rightarrow f_{6},f_{10}\rightarrow f_{7})$}\tabularnewline
\cline{2-8}
 & $\rho^{\pi}$ & $\frac{2}{3}d_{2}+4d_{3}$ & $\frac{2}{3}d_{2}+4d_{3}$ & $\frac{2}{3}d_{2}+4d_{3}$ & $-\frac{1}{2}d_{2}+3d_{3}$ & $3d_{3}$ & \tabularnewline
 & $\rho^{K}$ & $\frac{2}{3}d_{2}+4d_{3}$ & $-\frac{1}{3}d_{2}+4d_{3}$ & $-\frac{4}{3}d_{2}+4d_{3}$ & $\frac{1}{3}d_{2}+2d_{3}$ & $\frac{1}{3}d_{2}+2d_{3}$ & $-\frac{2}{3}d_{2}+8d_{3}$\tabularnewline
 & $\rho^{\eta}$ &  &  &  & $\frac{1}{2}d_{2}+3d_{3}$ & $-d_{2}+3d_{3}$ & \tabularnewline
\cline{2-8}
 & $\delta^{\pi}$ & $-1$ &  & $1$ & $\frac{1}{4}$ & $\frac{1}{2}$ & \tabularnewline
 & $\delta^{K}$ & $-1$ & $-\frac{1}{2}$ &  & $-\frac{1}{2}$ & $\frac{1}{2}$ & $1$\tabularnewline
 & $\delta^{\eta}$ &  &  &  & $-\frac{1}{4}$ &  & \tabularnewline
\midrule[1pt] \multirow{6}{*}{(k),(l)} & $\xi^{\pi}$ & $2$ & $2$ &
$2$ & $\frac{3}{4}$ & $\frac{3}{4}$ & \tabularnewline
 & $\xi^{K}$ & $1$ & $1$ & $1$ & $\frac{5}{2}$ & $\frac{5}{2}$ & $2$\tabularnewline
 & $\xi^{\eta}$ & $\frac{1}{3}$ & $\frac{1}{3}$ & $\frac{1}{3}$ & $\frac{1}{12}$ & $\frac{1}{12}$ & $\frac{4}{3}$\tabularnewline
\cline{2-8}
 & $\eta^{\pi}$ & $2$ & $2$ & $2$ & $\frac{3}{2}$ & $\frac{3}{2}$ & \tabularnewline
  & $\eta^{K}$ & $2$ & $2$ & $2$ & $1$ & $1$ & $4$\tabularnewline
 & $\eta^{\eta}$ &  &  &  & $\frac{3}{2}$ & $\frac{3}{2}$ & \tabularnewline
\bottomrule[1pt]\bottomrule[1pt]
\end{tabular}
\end{table}

\section{Independent LECs in the heavy quark limit}{\label{HQSS}}
There are eighteen unknown LECs in the analytical expressions in
Eqs.(\ref{mu1})-(\ref{mu3loop}), including five axial charges
$g_{1-5}$, ten $\mathcal{O}(p^2)$ baryon-photon coupling constants
$d_{2,3,5,6}$, $f_{2,4,6,7,9,10}$, one meson-meson-baryon coupling
constant $f_8$, and two $\mathcal{O}(p^4)$ chiral symmetry breaking
coupling constants $h_{2,4}$. Since the number of the LECs is larger
than that of the ground heavy baryons, we use the  heavy quark
symmetry to reduce the number of independent LECs.

The spin-$1\over 2$ and the spin-$3\over 2$ sextet are degenerate
states in the heavy quark limit. The heavy quark symmetry can relate
some LECs to others. We define a superfiled $\mathcal{H}_{\mu}$ to
denote $\mathcal{B}_6$ and
$\mathcal{B}_{6\mu}^*$~\cite{Cho:1992nt,Cheng:1993kp},
\begin{eqnarray}
&{\cal H}_{\mu}={\cal B}_{6\mu}^{*}-\sqrt{\frac{1}{3}}(\gamma_{\mu}+v_{\mu})\gamma^{5}{\cal B}_{6},\\
&\bar{{\cal H}}_{\mu}   =\bar{{\cal
B}}_{6\mu}^{*}+\sqrt{\frac{1}{3}}\bar{{\cal
B}}_{6}\gamma^{5}(\gamma_{\mu}+v_{\mu}),\label{superfield}
\end{eqnarray}
where $\bar{\mathcal{H}}_\mu$ is the conjugate field of
$\mathcal{H}_\mu$. $v_{\mu}$ is the velocity of heavy quark. In the
heavy quark limit, the $v_\mu$ also corresponds to the velocity of
the heavy baryon. This field $\mathcal{H}_\mu$ is constrained by
\begin{equation}
v\cdot\mathcal{H}=0,\quad \slashed{v}\mathcal{H}=\mathcal{H}.
\end{equation}
The $\mathcal{H}_\mu$ follows the same chiral transformation in
Eq.~(\ref{chiraltrans}).

In Refs.~\cite{Cho:1992nt,Cheng:1993kp}, the authors constructed the
axial coupling Lagrangian of the sextet baryons in heavy quark
symmetry,
\begin{equation}
{\cal L}_{{\cal
H}\phi}^{(1)}=ig_{a}\epsilon_{\mu\nu\rho\sigma}\langle\bar{{\cal
H}}^{\mu}u^{\rho}v^{\sigma}{\cal
H}^{\nu}\rangle+g_{b}\langle\bar{{\cal H}}^{\mu}u_{\mu}{\cal
B}_{\bar{3}}+{\rm H.c.}\rangle.\label{hq1Hphi}
\end{equation}
The LECs in $\mathcal{L}_{\mathcal{B}\phi}^{(1)}$ are reduced to two
independent LECs, $g_a$ and $g_b$:
\begin{equation}
g_{5}=g_{a},\quad g_{1}=-\frac{2}{3}g_{a},\quad
g_{3}=-\sqrt{\frac{1}{3}}g_{a},\quad g_{4}=g_{b};\quad
g_{2}=-\sqrt{\frac{1}{3}}g_{b},\quad g_6=0.
\end{equation}
$g_6$ is the coupling constant between pseudoscalar mesons and
antitriplet heavy baryons. The light spin $S_l=0$ for the
antitriplets. The pseudoscalar mesons only interact with the light
degree in the heavy baryon. Thus, the parity and angular momentum
conservation forbid the $g_6$ vertex.

The interaction in ${\cal L}_{\mathcal{B}\gamma}^{(2)}$ in heavy
quark symmetry corresponds to
\begin{equation}
{\cal L}_{{\cal
H}\gamma}^{(2)}=i\frac{g_{c}}{4M_{N}}\langle\bar{{\cal
H}}^{\mu}\hat{F}_{\mu\nu}^{+}{\cal
H}^{\nu}\rangle-\frac{g_{e}}{4M_{N}}\epsilon^{\sigma\mu\nu\rho}\langle\bar{{\cal
H}}_{\sigma}\hat{F}_{\mu\nu}^{+}v_{\rho}{\cal B}_{3}\rangle
+\text{H.c.}
\end{equation}
The eight LECs $d_{5}$, $d_6$, $f_2$, $f_4$, $f_6$, $f_7$, $f_9$ and
$f_{10}$ in ${\cal L}_{B\gamma}^{(2)}$ are reduced to two LECs $g_c$
and $g_e$:
\begin{eqnarray}
&f_{9}=g_{c},\quad f_{6}=\frac{2}{\sqrt{3}}g_{c},\quad d_{5}=-\frac{2}{3}g_{c},\nonumber \\
&f_{10}=f_{7}= d_{6}=0, \nonumber \\
&f_{2}=-\frac{2}{\sqrt{3}}g_{e},\quad f_{4}=-4g_{e}. \label{gcde}
\end{eqnarray}

In Lagrangians, we decompose the $F^+_{\mu\nu}$ into the trace part
$\langle F^+_{\mu\nu}\rangle$ and traceless part
$\hat{F}^+_{\mu\nu}$, which correspond to the contributions from the
light quarks and the heavy quark, respectively. The contribution
from the heavy quark to the magnetic moments is order of $1\over
m_c$. Thus, heavy quark contribution and the LECs, $f_{10}$, $f_7$
and $d_6$ vanish in the heavy quark limit.

For spin-$1\over 2$ singly heavy baryons, we construct the
$\mathcal{O}(p^2)$ meson-baryon interaction and $\mathcal{O}(p^4)$
photon-baryon interaction~\cite{Wang:2018gpl}:
\begin{eqnarray}
&\mathcal{L}_{B\phi\phi}^{(2)}=\frac{id_{4}}{2M_{N}}\langle\bar{B}_{6}\sigma^{\mu\nu}[u_{\mu},u_{\nu}]B_{6}\rangle,\label{spin1over2a}\\
&\mathcal{L}_{B\gamma}^{(4)}=\frac{s_{2}}{8M_N}\langle\bar{B}_{6}\sigma^{\mu\nu}\langle
F_{\mu\nu}^{+}\rangle{\tilde{\chi}}_{+}B_{6}\rangle+\frac{s_{4}}{2M_N}\langle\bar{B}_{6}\sigma^{\mu\nu}F_{\mu\nu}^{+}B_{6}{\tilde{\chi}}^{T}_+\rangle.\label{spin1over2}
\end{eqnarray}
In the heavy quark symmetry, the LECs in Eqs.~(\ref{spin1over2a})
and (\ref{spin1over2}) can be related to those in
Eqs.~(\ref{2Bphiphi}) and (\ref{4Bgamma}). The Lagrangians in the
heavy quark limit read
\begin{eqnarray}
&{\cal L}_{{\cal H}\phi\phi}^{(2)}=\frac{g_{f}}{4M_{N}}\langle\bar{{\cal H}}^{\mu}[u_{\mu},u_{\nu}]{\cal H}^{\nu}\rangle \nonumber \\
&{\cal L}_{{\cal
H}\gamma}^{(4)}=i\frac{g_{h}}{4M_{N}}\langle\bar{{\cal
H}}^{\mu}\hat{F}_{\mu\nu}^{+}{\cal
H}^{\nu}{\tilde{\chi}}^{T}_+\rangle
\end{eqnarray}
The LECs are related as
\begin{eqnarray}
d_{4}=\frac{1}{6}g_{f};\quad f_{8}=-\frac{1}{2}g_{f},\quad
s_{2}=h_{2}=0;\quad s_{4}=-\frac{1}{6}g_{h};\quad
h_{4}=\frac{g_{h}}{4} .\label{ggfh}
\end{eqnarray}

In the heavy quark limit, the heavy quark contribution vanishes.
Thus, in this limit, there are seven nonvanishing independent LECs,
$d_{2}$ and $g_{a,b,c,e,f,h}$, contributing to the magnetic moments
of the sextet baryons.

\section{numerical results}\label{nrslt}
In the present work, we perform the numerical analysis with three
scenarios. In the first scenario, we use the LECs determined by our
previous work~\cite{Wang:2018gpl}. Three new LECs, $f_8$, $h_2$ and
$h_4$ are related to $d_4$, $s_2$ and $s_4$ through the heavy quark
spin symmetry. In the second scenario, we reduce the number of the
LECs in the heavy quark limit and adopt the lattice QCD simulation
results as input. In the third scenario, we include the heavy quark
contribution on the basis of the scenario II. As a by-product, we
also give the magnetic moments of singly bottom baryon.

In the three scenarios, we all use the same axial coupling values.
The axial coupling constants $g_2$ and $g_4$ in Eq.~(\ref{1Bphi})
are estimated through the decay widths of $\Sigma_c$ and
$\Sigma_c^*$, respectively~\cite{Meguro:2011nr,Patrignani:2016xqp}.
The other $g_i$ are related to $g_2$ and $g_4$ with the help of
quark model. Their values are
\begin{eqnarray}
&g_2=-0.60,\quad g_4=-\sqrt{3}g_2=1.04,\quad g_1=-\sqrt{8\over 3}g_2=0.98,\quad \nonumber\\
&g_3=\frac{\sqrt{3}}{2}g_1=0.85, \quad g_5=-\frac{3}{2}g_1=-1.47,
\quad g_6=0.\label{cpaxial}
\end{eqnarray}

\begin{table}
\caption{The lattice QCD simulation
results~\cite{Can:2013tna,Bahtiyar:2016dom,Can:2015exa}. ``$\surd$"
represents the results used as input. }\label{LQCD}
\begin{tabular}{l|cc|ccccc|c}
\toprule[1pt]\toprule[1pt]
 & $\Xi_{c}^{+}$ & $\Xi_{c}^{0}$ & $\Sigma_{c}^{++}$ & $\Sigma_{c}^{0}$ & $\Xi_{c}^{'+}$ & $\Xi_{c}^{'0}$ & $\Omega_{c}^{0}$ & $\Omega_{c}^{*0}$\tabularnewline
\midrule[1pt] LQCD & 0.235(25) & 0.192(17) & 1.499(202) &
-0.875(103) & 0.315(141) & -0.599(71) & -0.688(31) &
-0.730(23)\tabularnewline SI Input & $\surd$ & $\surd$ & $\surd$ & &
$\surd$ &  & $\surd$ & \tabularnewline SII Input & $\surd$ & $\surd$
& $\surd$ &  & $\surd$ &  & $\surd$ & $\surd$\tabularnewline SIII
Input & $\surd$ & $\surd$ & $\surd$ &  & $\surd$ &  & $\surd$ &
$\surd$\tabularnewline \bottomrule[1pt]\bottomrule[1pt]
\end{tabular}
\end{table}

\subsection{Scenario I}

In our previous work~\cite{Wang:2018gpl}, we calculated the magnetic
moments of the spin-$1\over 2$ singly heavy baryons. All the LECs in
Eq. (\ref{2Bgamma}) have been evaluated through the quark model.
Here, we review the idea in brief. The vertices in
Eq.~(\ref{2Bgamma}) contribute to the leading order (transition)
magnetic moments in the HBChPT scheme. We assume that their values
are approximate to those estimated by the naive quark model. Then,
we can extract these LECs. The (transition) magnetic moments from
the quark model and the leading order results in HBChPT are given in
Table~\ref{qm3} and \ref{qm6}.

\begin{table}
\caption{The (transition) magnetic moments $\mu_{B_{\bar{3}}}$,
$\mu_{B_{6}\rightarrow B_{\bar{3}}}$, and $\mu_{B_{6}^{*}\rightarrow
B_{\bar{3}}}$ from the quark model and the leading order results in
HBChPT.}\label{qm3}
\begin{tabular}{c|c|c|c}
\toprule[1pt] \toprule[1pt] $\mu_{B_{\bar{3}}}$ & $\Lambda_{c}^{+}$
& $\Xi_{c}^{+}$ & $\Xi_{c}^{0}$\tabularnewline \midrule[1pt] QM &
$\mu_{c}$ & $\mu_{c}$ & $\mu_{c}$\tabularnewline ${\cal O}(p^{1})$ &
$\frac{1}{3}d_{2}+2d_{3}$ & $\frac{1}{3}d_{2}+2d_{3}$ &
$-\frac{2}{3}d_{2}+2d_{3}$\tabularnewline \midrule[1pt]
$\mu_{B_{6}\rightarrow B_{\bar{3}}}$ &
$\Sigma_{c}^{+}\rightarrow\Lambda_{c}^{+}\gamma$ &
$\Xi_{c}^{'+}\rightarrow\Xi_{c}^{+}\gamma$ &
$\Xi_{c}^{'0}\rightarrow\Xi_{c}^{0}\gamma$\tabularnewline
\midrule[1pt] QM & $\sqrt{\frac{1}{3}}(\mu_{d}-\mu_{u})$ &
$\sqrt{\frac{1}{3}}(\mu_{s}-\mu_{u})$ &
$\sqrt{\frac{1}{3}}(\mu_{s}-\mu_{d})$\tabularnewline ${\cal
O}(p^{1})$ & $\sqrt{\frac{1}{2}}f_{2}$ & $\sqrt{\frac{1}{2}}f_{2}$ &
0\tabularnewline \midrule[1pt] $\mu_{B_{6}^{*}\rightarrow
B_{\bar{3}}}$ & $\Sigma_{c}^{*+}\rightarrow\Lambda_{c}^{+}\gamma$ &
$\Xi_{c}^{*+}\rightarrow\Xi_{c}^{+}\gamma$ &
$\Xi_{c}^{*0}\rightarrow\Xi_{c}^{0}\gamma$\tabularnewline
\midrule[1pt] QM & $\frac{2}{\sqrt{6}}(\mu_{u}-\mu_{d})$ &
$\frac{2}{\sqrt{6}}(\mu_{u}-\mu_{s})$ &
$\frac{2}{\sqrt{6}}(\mu_{d}-\mu_{s})$\tabularnewline ${\cal
O}(p^{1})$ & $-\sqrt{\frac{1}{12}}f_{4}$ &
$-\sqrt{\frac{1}{12}}f_{4}$ & 0\tabularnewline
\bottomrule[1pt]\bottomrule[1pt]
\end{tabular}
\end{table}

\begin{table}
\caption{The (transition) magnetic moments $\mu_{B_{\bar{6}}}$,
$\mu_{B^*_{\bar{6}}}$ and $\mu_{B_{6}^{*}\rightarrow B_{6}}$ from
the quark model and the leading order results in HBChPT.}\label{qm6}
\begin{tabular}{c|c|c|c|c|c|c}
\toprule[1pt]\toprule[1pt] $\mu_{B_{6}}$ & $\Sigma_{c}^{++}$ &
$\Sigma_{c}^{+}$ & $\Sigma_{c}^{0}$ & $\Xi_{c}^{\prime+}$ &
$\Xi_{c}^{\prime 0}$ & $\Omega_{c}^{0}$\tabularnewline \midrule[1pt]
QM & $\frac{4}{3}\mu_{u}-\frac{1}{3}\mu_{c}$ &
$\frac{2}{3}\mu_{u}+\frac{2}{3}\mu_{d}-\frac{1}{3}\mu_{c}$ &
$\frac{4}{3}\mu_{d}-\frac{1}{3}\mu_{c}$ &
$\frac{2}{3}\mu_{u}+\frac{2}{3}\mu_{s}-\frac{1}{3}\mu_{c}$ &
$\frac{2}{3}\mu_{d}+\frac{2}{3}\mu_{s}-\frac{1}{3}\mu_{c}$ &
$\frac{4}{3}\mu_{s}-\frac{1}{3}\mu_{c}$\tabularnewline ${\cal
O}(p^{1})$ & $\frac{2}{3}d_{5}+d_{6}$ & $\frac{1}{6}d_{5}+d_{6}$ &
$-\frac{1}{3}d_{5}+d_{6}$ & $\frac{1}{6}d_{5}+d_{6}$ &
$-\frac{1}{3}d_{5}+d_{6}$ & $-\frac{1}{3}d_{5}+d_{6}$\tabularnewline
\midrule[1pt] $\mu_{B_{6}^{*}}$ & $\Sigma_{c}^{*++}$ &
$\Sigma_{c}^{*+}$ & $\Sigma_{c}^{*0}$ & $\Xi{}_{c}^{*+}$ &
$\Xi{}_{c}^{*0}$ & $\Omega_{c}^{*0}$ \tabularnewline \midrule[1pt]
QM & $2\mu_{u}+\mu_{c}$ & $\mu_{u}+\mu_{d}+\mu_{c}$ &
$2\mu_{d}+\mu_{c}$ & $\mu_{u}+\mu_{s}+\mu_{c}$ &
$\mu_{d}+\mu_{s}+\mu_{c}$ & $2\mu_{s}+\mu_{c}$\tabularnewline ${\cal
O}(p^{1})$ & $-\frac{2}{3}f_{9}-f_{10}$ & $-\frac{1}{6}f_{9}-f_{10}$
& $\frac{1}{3}f_{9}-f_{10}$ & $-\frac{1}{6}f_{9}-f_{10}$ &
$\frac{1}{3}f_{9}-f_{10}$ & $\frac{1}{3}f_{9}-f_{10}$\tabularnewline
\midrule[1pt] $\mu_{B_{6}^{*}\rightarrow B_{6}}$ &
$\Sigma_{c}^{*++}\rightarrow\Sigma_{c}^{++}\gamma$ &
$\Sigma_{c}^{*+}\rightarrow\Sigma_{c}^{+}\gamma$ &
$\Sigma_{c}^{*0}\rightarrow\Sigma_{c}^{0}\gamma$ &
$\Xi{}_{c}^{*+}\rightarrow\Xi_{c}^{\prime+}\gamma$ &
$\Xi{}_{c}^{*0}\rightarrow\Xi_{c}^{\prime0}\gamma$ &
$\Omega{}_{c}^{*0}\rightarrow\Omega^{0}\gamma$\tabularnewline
\midrule[1pt] QM & $\frac{2\sqrt{2}}{3}(\mu_{u}-\mu_{c})$ &
$\frac{\sqrt{2}}{3}(\mu_{u}+\mu_{d}-2\mu_{c})$ &
$\frac{2\sqrt{2}}{3}(\mu_{d}-\mu_{c})$ &
$\frac{\sqrt{2}}{3}(\mu_{u}+\mu_{s}-2\mu_{c})$ &
$\frac{\sqrt{2}}{3}(\mu_{d}+\mu_{s}-2\mu_{c})$ &
$\frac{2\sqrt{2}}{3}(\mu_{s}-\mu_{c})$\tabularnewline ${\cal
O}(p^{1})$ & $-\sqrt{\frac{1}{6}}(\frac{2}{3}f_{6}+f_{7})$ &
$-\sqrt{\frac{1}{6}}\left(\frac{1}{6}f_{6}+f_{7}\right)$ &
$-\sqrt{\frac{1}{6}}\left(-\frac{1}{3}f_{6}+f_{7}\right)$ &
$-\sqrt{\frac{1}{6}}\left(\frac{1}{6}f_{6}+f_{7}\right)$ &
$-\sqrt{\frac{1}{6}}\left(-\frac{1}{3}f_{6}+f_{7}\right)$ &
$-\sqrt{\frac{1}{6}}\left(-\frac{1}{3}f_{6}+f_{7}\right)$\tabularnewline
\bottomrule[1pt]\bottomrule[1pt]
\end{tabular}
\end{table}

Apart from the axial coupling constants and the $\mathcal{O}(p^2)$
baryon-photon coupling constants, we have three new LECs in the
present work. In our previous work, $d_4$, $s_2$ and $s_4$ have been
determined by fitting three lattice QCD results,
$\mu_{\Sigma_{c}^{++}}$, $\mu_{\Xi_{c}^{\prime+}}$, and
$\mu_{\Omega_{c}^{0}}$ in Table~\ref{LQCD}. With the relations in
Eq.(\ref{ggfh}), we obtain the values of $f_8$, $h_2$ and $h_4$. In
this scenario, we keep the mass splitting between the spin-$1\over
2$ sextet and the spin-$3\over 2$ sextet. The mass splitting reads
\begin{eqnarray}
\delta_{1}  &=&M_{6*}-M_{6}=67~\text{MeV},\nonumber\\
\delta_{2}  &=&M_{6}-M_{\bar{3}}=127~\text{MeV},\nonumber\\
\delta_{3}  &=&M_{6*}-M_{\bar{3}}=194~\text{MeV}.
\end{eqnarray}

We have determined all the LECs in the analytical expressions up to
$\mathcal{O}(p^3)$. We give the numerical results in two schemes. In
the first scheme, we include the spin-$1\over 2$ antitriplet, the
spin-$1\over 2$, and the spin-$3\over 2$ sextet as the intermediate
states in the loops. The numerical results are listed in the left
panel of Table~\ref{SI}. The chiral convergence is not good enough.
The $\mathcal{O}(p^3)$ contribution is larger than that at
$\mathcal{O}(p^2)$ for $\Xi_c^{*+}$ and $\Sigma_c^{*0}$. In the
second scheme, we only take the spin-$1\over 2$ and spin-$3\over 2$
sextet as the intermediate states. The results are given in the
right panel of Table~\ref{SI}. The chiral convergence becomes much
better.

The mass splittings $\delta_{1,2,3}$ do not vanish in the chiral
limit, which will worsen the chiral convergence. Due to the large
mass splitting $\delta_3$, about 194 MeV, including the spin-$1\over
2$ antitriplet will destroy the chiral convergence.  As for the
spin-$1\over 2$ sextet, the mass splitting $\delta_1$ is small.
Taking  spin-$1\over 2$ sextet as intermediate states has almost no
negative impact on the convergence. Meanwhile, the spin-$1\over 2$
and spin-$3\over 2$ sextet form doublet in the heavy quark limit. To
calculate  the magnetic moments of the spin-$3\over 2$ sextet, the
contribution from the chiral fluctuation around the spin-$1\over 2$
sextet is important. Thus, we choose the results from the second
scheme, in the right panel of Table~\ref{SI}, as our final results.
\begin{table}
\caption{The numerical results of spin-$3\over 2$ singly charmed
baryon magnetic moments in the scenario I (in unit of $\mu_N$). We
take $B_{\bar{3}}$, $B_6$ and $B^*_6$ as the intermediate states in
the left panel, while we only consider the $B_6$ and $B^*_6$ in the
right panel. }\label{SI}
\begin{tabular}{l|cccc|cccc}
\toprule[1pt]\toprule[1pt] SI  & \multicolumn{4}{c|}{with
$B_{\bar{3}}$, $B_6$ and $B^*_{6}$} & \multicolumn{4}{c}{with $B_6$
and $B^*_{6}$}\tabularnewline \hline
 & ${\cal O}(p^{1})$ & ${\cal O}(p^{2})$ & ${\cal O}(p^{3})$ & Total & ${\cal O}(p^{1})$ & ${\cal O}(p^{2})$ & ${\cal O}(p^{3})$ & Total\tabularnewline
\midrule[1pt] $\Sigma_{c}^{*++}$ & 4.10 & -1.16 & -0.02 & 2.92 &
4.10 & -1.03 & -0.16 & 2.91\tabularnewline $\Sigma_{c}^{*+}$ & 1.48
& -0.72 & -0.17 & 0.59  & 1.48 & -0.39 & -0.10 & 0.99\tabularnewline
$\Sigma_{c}^{*0}$ & -1.13 & -0.29 & -0.32 & -1.74& -1.13 & 0.26 &
-0.05 & -0.92\tabularnewline $\Xi_{c}^{*+}$ & 1.48     & 0.14 &
-0.50 & 1.13      & 1.48 & -0.13 & -0.07 & 1.28\tabularnewline
$\Xi_{c}^{*0}$ & -1.13    & 0.58 & -0.22 & -0.77    & -1.13 & 0.52 &
0.003 & -0.61\tabularnewline $\Omega_{c}^{*0}$ & -1.13 & 1.45 &
-0.27 & 0.05  & -1.13 & 0.78 & 0.08 & -0.27\tabularnewline
\bottomrule[1pt]\bottomrule[1pt]
\end{tabular}
\end{table}

\subsection{Scenario \rmII}
According to Section~\ref{HQSS}, we reduce the LECs to five unknown
independent ones with the heavy quark symmetry. In this scenario, we
make use of six lattice QCD results in Table~\ref{LQCD} to determine
them. In the heavy quark limit, the spin-$1\over 2$ and $3\over 2$
are degenerate states. The mass splittings are
\begin{eqnarray}
\delta_{1}  &=&M_{6^*}-M_{6}=0~\text{MeV}, \nonumber \\
\delta_{2}  &=&\delta_{3}=M_{6^{(*)}}-M_{\bar{3}}=161~\text{MeV}.
\end{eqnarray}
In this scenario, we also apply two schemes to estimate the LECs. In
the first scheme, we consider all the singly charmed baryons as the
intermediate states. The results are given in the left panel of
Table~\ref{SII}. In the second scheme, we set $g_e=0$ and decouple
the $\bar{3}_f$ and $6_f$ singly charmed baryon in the loop
diagrams. The results are given in the right panel of
Table~\ref{SII}.

The results in the left panel suffer from the bad convergence, which
are even worse than those in the first scheme in the scenario I. The
quark model predictions are comparable with the lattice QCD results.
Taking the quark model results as the leading order input at least
ensure a dominant $\mathcal{O}(p^1)$ contribution in scenario I.
Comparing results in the two panels of Table~\ref{SII}, although the
total values are similar, the chiral convergence in the second
scheme improve significantly. Including the antitriplet as the
intermediate states break the chiral convergence. Thus, we also
choose the results from the second scheme as our final results in
this scenario.

In the second scheme, the magnetic moments of sextet baryons do not
depend on the antitriplet.  In the sextet sector, we determine three
unknown LECs and obtain twelve magnetic moments. Thus, this scenario
has powerful predictions.

 \begin{table}
 \caption{The numerical results of spin-$3\over 2$ singly charmed baryon magnetic moments in the scenario II (in unit of $\mu_N$). We take $B_{\bar{3}}$, $B_6$ and $B^*_6$ as the intermediate states in the left panel, while we only consider the $B_6$ and $B^*_6$ in the right panel.}\label{SII}
 \begin{tabular}{l|cccc|cccc}
\toprule[1pt]\toprule[1pt] SII  & \multicolumn{4}{c|}{with
$B_{\bar{3}}$, $B_6$ and $B^*_{6}$} & \multicolumn{4}{c}{with $B_6$
and $B^*_{6}$}\tabularnewline \hline
 & ${\cal O}(p^{1})$ & ${\cal O}(p^{2})$ & ${\cal O}(p^{3})$ & Total & ${\cal O}(p^{1})$ & ${\cal O}(p^{2})$ & ${\cal O}(p^{3})$ & Total\tabularnewline
\midrule[1pt] $\Sigma_{c}^{*++}$ & 0.78 & -1.28 & 2.43 & 1.92 & 2.63
& -1.11 & 0.60 & 2.12\tabularnewline $\Sigma_{c}^{*+}$  & 0.19 &
-0.74 & 0.82 & 0.27  & 0.66 & -0.39 & 0.18 & 0.44\tabularnewline
$\Sigma_{c}^{*0}$  & -0.39 & -0.20 & -0.78 &-1.37& -1.32 & 0.32 &
-0.25 & -1.24\tabularnewline $\Xi_{c}^{*+}$     & 0.19 & 0.10 &
-0.03 & 0.27     & 0.66 & -0.16 & 0.03 & 0.52\tabularnewline
$\Xi_{c}^{*0}$     & -0.39 & 0.64 & -1.28 & -1.03   & -1.32 & 0.55 &
-0.27 & -1.03\tabularnewline $\Omega_{c}^{*0}$  & -0.39 & 1.49 &
-1.83 & -0.73& -1.32 & 0.78 & -0.26 & -0.79\tabularnewline
\bottomrule[1pt]\bottomrule[1pt]
\end{tabular}
 \end{table}

 \subsection{Scenario \rmIII}
In the lattice QCD
simulation~\cite{Can:2013tna,Bahtiyar:2016dom,Can:2015exa}, the
contribution of heavy quark and light quarks to the magnetic moments
are given separately. The heavy quark contribution for $\Xi_c^+$,
$\Sigma^{++}$, $\Xi_c^{\prime +}$, $\Omega_c^0$ and $\Omega_c^{*0}$
read
\begin{eqnarray}
\mu^{c}_{\Xi_c^+}=0.226\mu_N,\quad
\mu^c_{\Sigma_c^{++}}=-0.066\mu_N\quad
\mu^c_{\Xi_c^{\prime+}}=-0.059\mu_N,\quad
\mu^c_{\Omega_c^0}=-0.061\mu_N,\quad
\mu^c_{\Omega_c^{*0}}=0.239\mu_N,\label{hqlqcd}
\end{eqnarray}
where the superscript ``$c$" denotes the contribution from the charm
quark. According to the quark model in Tables \ref{qm3} and
\ref{qm6}, the heavy quark contribution is $\mu_c$, $-{1\over
3}\mu_c$ and $\mu_c$ for the antitriplet, spin-$1\over 2$ sextet and
spin-$3\over 2$ sextet, respectively. Using the lattice QCD results
in Eqs.~(\ref{hqlqcd}), we get the average $\mu_c=0.205\mu_N $. In
this scenario, the heavy quark contribution is estimated by using
the average $\mu_c$ while the light quark contribution is derived
through fitting the remaining part of lattice QCD results. The
results are given in Table~\ref{SIII}. The right panel of this table
is our final results of this scenario.

The heavy quark contribution can also be introduced through the
heavy quark symmetry breaking Lagrangian at $\mathcal{O}(1/m_c)$
which reads,
\begin{equation}
{\cal L}_{HQ}={g_H\over 8M_N}\langle\bar{{\cal
H}}^{\rho}\sigma_{\mu\nu}{\cal H}_{\rho}\rangle\langle
F^{+\mu\nu}\rangle
\end{equation}
The LECs $d_6$, $f_7$ and $f_{10}$ are related to $g_H$ as
\begin{equation}
f_{10}={g_H},\quad d_6=\frac{1}{3}{g_H},\quad
f_{7}=-\frac{4}{\sqrt{3}}{g_H}.
\end{equation}
We use the heavy quark contribution from the lattice QCD simulation
to extract the $g_H$. The same magnetic moment results are obtained.

In scenario \rmIII, we can easily extend our formalism to calculate
the magnetic moments of singly bottom baryons. In the heavy quark
limit, the light contribution for a bottom baryon is the same as
that for the charmed baryon. The heavy quark part is estimated using
the quark model. We adopt the constituent mass $m_b=4700$ MeV. The
magnetic moments  of singly botom baryons are given in Table
\ref{SIIIb}.

\begin{table}
\caption{The numerical results of spin-$3\over 2$ singly charmed
baryon magnetic moments from the scenario III (in unit of $\mu_N$).
We take $B_{\bar{3}}$, $B_6$ and $B^*_6$ as the intermediate states
in the left panel, while we only take the $B_6$ and $B^*_6$ in the
right panel. The ``HQ" represents the heavy quark
contribution.}\label{SIII}
\begin{tabular}{l|ccccc|ccccc}
\toprule[1pt]\toprule[1pt] SIII &  & \multicolumn{4}{c|}{with
$B_{\bar{3}}$, $B_6$ and $B^*_{6}$} &  & \multicolumn{4}{c}{with
$B_6$ and $B^*_{6}$}\tabularnewline \hline
 & HQ & ${\cal O}(p^{1})$ & ${\cal O}(p^{2})$ & ${\cal O}(p^{3})$ & Total & HQ & ${\cal O}(p^{1})$ & ${\cal O}(p^{2})$ & ${\cal O}(p^{3})$ & Total\tabularnewline
\midrule[1pt] $\Sigma_{c}^{*++}$ & 0.21 & 0.96 & -1.28 & 2.65 & 2.54
& 0.21 & 2.69 & -1.11 & 0.62 & 2.41\tabularnewline $\Sigma_{c}^{*+}$
& 0.21  & 0.24 & -0.74 & 0.96 & 0.66 & 0.21 & 0.67 & -0.39 & 0.18 &
0.67\tabularnewline $\Sigma_{c}^{*0}$ & 0.21 & -0.48 & -0.20 & -0.73
& -1.21& 0.21 & -1.35 & 0.32 & -0.26 & -1.07\tabularnewline
$\Xi_{c}^{*+}$ & 0.21 &  0.24    & 0.10 & 0.27 & 0.81   & 0.21 &
0.67 & -0.16 & 0.10 & 0.81\tabularnewline $\Xi_{c}^{*0}$ & 0.21 &
-0.48    & 0.64 & -1.31 & -0.94 & 0.21 & -1.35 & 0.55 & -0.31 &
-0.90\tabularnewline $\Omega_{c}^{*0}$ & 0.21 & -0.48 & 1.49 & -1.94
& -0.73 & 0.21 & -1.35 & 0.78 & -0.34 & -0.70\tabularnewline
\bottomrule[1pt]\bottomrule[1pt]
\end{tabular}
\end{table}

\begin{table}
\caption{The magnetic moments of singly bottom baryon sextet (in
unit of $\mu_N$). The ``HQ" represents the heavy quark contribution.
The light quark contritbution is the same as that for singly charmed
baryon. }\label{SIIIb}
\begin{tabular}{lcc|ccc}
\toprule[1pt]\toprule[1pt] spin-$\frac{1}{2}$ & HQ & Total &
spin-$\frac{3}{2}$ & HQ & Total\tabularnewline \midrule[1pt]
$\Sigma_{b}^{+}$ & 0.02 & 1.59 & $\Sigma_{b}^{*+}$ & -0.06 &
2.14\tabularnewline $\Sigma_{b}^{0}$ & 0.02 & 0.39 &
$\Sigma_{b}^{*0}$ & -0.06 & 0.40\tabularnewline $\Sigma_{b}^{-}$ &
0.02 & -0.81 & $\Sigma_{b}^{*-}$ & -0.06 & -1.35\tabularnewline
$\Xi_{b}^{0}$ & 0.02 & 0.40 & $\Xi_{b}^{*0}$ & -0.06 &
0.54\tabularnewline $\Xi_{b}^{-}$ & 0.02 & -0.73 & $\Xi_{b}^{*-}$ &
-0.06 & -1.17\tabularnewline $\Omega_{b}^{-}$ & 0.02 & -0.65 &
$\Omega_{b}^{*-}$ & -0.06 & -0.97\tabularnewline \bottomrule[1pt]
\bottomrule[1pt]
\end{tabular}
\end{table}

\begin{table}
\caption{The numerical results of LECs for the three
scenarios.}\label{lecno}
\begin{tabular}{c|c|c|ccc|ccc|cc|cc|cc}
\toprule[1pt]\toprule[1pt] SI & $d_{2}$ & $d_{3}$ & $f_{9}$ &
$f_{6}$ & $d_{5}$ & $f_{10}$ & $f_{7}$ & $d_{6}$  & $d_{4}$ &
$f_{8}$ & $s_{2}$ & $h_{2}$ & $s_{4}$ & $h_{4}$\tabularnewline
\hline
 & 0.04 & 0.11 & -5.23 & -6.00 & 3.49 & -0.61 & 0.60 & 0.03 &  3.45 & -10.35 & -0.24 & 0.36 & -0.04 & 0.05\tabularnewline
\midrule[1pt] SII & $d_{2}$ & $d_{3}$ & \multicolumn{3}{c|}{$g_{c}$}
& $f_{10}$ & $f_{7}$ & $d_{6}$ & \multicolumn{2}{c|}{$g_{f}$} &
$s_{2}$ & $h_{2}$ & \multicolumn{2}{c}{$g_{h}$}\tabularnewline
\hline
 & -0.09 & 0 & \multicolumn{3}{c|}{-3.95} & \multicolumn{3}{c|}{0} & \multicolumn{2}{c|}{0.61} & \multicolumn{2}{c|}{0} & \multicolumn{2}{c}{0.32}\tabularnewline
\midrule[1pt] SIII & $d_{2}$ & $d_{3}$ &
\multicolumn{3}{c|}{$g_{c}$} & $f_{10}$ & $f_{7}$ & $d_{6}$ &
\multicolumn{2}{c|}{$g_{f}$} & $s_{2}$ & $h_{2}$ &
\multicolumn{2}{c}{$g_{h}$}\tabularnewline \hline
 & 0.03 & 0.10 & \multicolumn{3}{c|}{-4.04} &-0.21&0.47&-0.07 & \multicolumn{2}{c|}{0.41} & \multicolumn{2}{c|}{0} & \multicolumn{2}{c}{0.12}\tabularnewline
\bottomrule[1pt]\bottomrule[1pt]
\end{tabular}
\end{table}

\section{Discussion and Conculsion}\label{Disscss}
\begin{table}
\caption{Comparision of the spin-$3\over 2$ singly charmed baryon
magnetic in the literature, including the lattice QCD
(LQCD)~\cite{Can:2015exa}, the hyper central model
(HCM)~\cite{Patel:2007gx}, effective mass (EM) and screened charge
scheme (SC)~\cite{Dhir:2009ax}, chiral constituent quark model
($\chi$CQM)~\cite{Sharma:2010vv}, light-cone QCD sum rules
(LCQSR)~\cite{Aliev:2008sk}, MIT bag
model~\cite{Bose:1980vy,Bernotas:2012nz}, Skyrmion~\cite{Oh:1991ws}
scheme and chiral quark-soliton model
($\chi$QSM)~\cite{Yang:2018uoj} (in unit of $\mu_N$).
}\label{rsltcomp}
\begin{tabular}{l|ccc|cccccccccc}
\toprule[1pt]\toprule[1pt]
 & SI & SII &SIII& LQCD& HCM & EM & SC & $\chi$CQM & LCQSR & Bag I & Bag \rmII & Skyrmion &$\chi$QSM\tabularnewline
\midrule[1pt] $\Sigma_{c}^{*++}$ & 2.91 & 2.12 &2.41 & - &
$3.68\sim3.84$ & 3.56 & 3.63 & 3.92 & $4.81\pm1.22$ & 3.91 & 3.13 &
$4.52\sim4.58$& $3.22\pm0.15$\tabularnewline
 $\Sigma_{c}^{*+}$ & 0.99 & 0.44 &0.67& - & $1.20\sim1.26$ & 1.17 & 1.18 & 0.97 &
$2.00\pm0.46$ & 1.34 & 1.09 & $1.12\sim1.31$ &
$0.68\pm0.04$\tabularnewline $\Sigma_{c}^{*0}$ & -0.92 & -1.24
&-1.07& - & $-0.83\sim-0.85$ & -1.23 & -1.18 &  -1.99 &
$-0.81\pm0.20$ & -1.20 & -0.96 & $-2.29\sim-1.92$ &
$-1.86\pm0.07$\tabularnewline $\Xi_{c}^{*+}$ & 1.28 & 0.52 & 0.81& -
& $1.45\sim1.52$ & 1.43 & 1.39 & 1.59 & $1.68\pm0.42$ & 1.54 & 1.27
& $2.26\sim2.07$ & $0.90\pm0.04$\tabularnewline
 $\Xi_{c}^{*0}$ &-0.61 & -1.03& -0.90& - & $-0.67\sim-0.69$ & -1.00 & -1.02 & -1.43 &
$-0.68\pm0.18$ & -1.01 & -0.75 & $-2.01\sim-1.98$ &
$-1.57\pm0.06$\tabularnewline $\Omega_{c}^{*0}$ & -0.27 & -0.79
&-0.70& -0.73 & $-0.83\sim-0.87$ & -0.77 & -0.84 & -0.86 &
$-0.62\pm0.18$ & -0.78 & -0.55 & $-0.87\sim-1.23$ &
$-1.28\pm0.08$\tabularnewline \bottomrule[1pt]\bottomrule[1pt]
\end{tabular}
\end{table}

We calculate the magnetic moments of spin-$3\over 2$ singly charmed
baryons. The analytical expressions are derived up to
$\mathcal{O}(p^3)$. There are eighteen unknown LECs involved. We
reduce them into seven novanishing independent LECs with the heavy
quark symmetry. Our numerical results are given up to
$\mathcal{O}(p^3)$ in three scenarios. In the first scenario, we
keep the mass difference between spin-$1\over 2$ and spin-$3\over 2$
sextet. The quark model results are regarded as the leading order
magnetic moments. Five lattice QCD results are used to determine the
LECs. The heavy quark symmetry is used to relate the
$\mathcal{O}(p^2)$ $B\phi\phi$ and $\mathcal{O}(p^4)$ $B\gamma$
vertices to those for the spin-$1\over 2$ heavy baryons. In the
second scenario, we adopt the heavy quark symmetry globally. The
spin-$1 \over 2$ and spin-$3\over 2$ sextets belong to the same
doublet. The five unknown LECs are fitted using six lattice QCD
results. In the third scenario, we add the heavy quark contribution
explicitly on the basis of scenario II. In this scenario, we also
evaluate the magnetic moments of singly bottom baryons as a
by-product. Including the spin-$1\over 2$ antitriplet intermediate
states will worsen the chiral divergence, due to its large mass
difference with the sextet. We list both the results with all
intermediate states and only sextet intermediate states.  We take
the latter ones as the final results.

We give our final results and those from other schemes in
Table~\ref{rsltcomp}.  Compared with the scenario II, the scenario
III includes the heavy quark contribution. The results in scenario
III tend to be closer to those from other schemes. Thus, the $1\over
m_c$ effect may be not negligible. While the bottom quark is much
heavier, its contribution in the singly bottom baryons can be
neglected. In the scenario I, no lattice QCD results for
spin-$3\over 2$ heavy baryon is used as input. The value of
$\mu_{\Omega_c^{*0}}$  in scenario I may become larger if we use
lattice QCD simulation value  as input. In the scenario III, we
determined three unknown LECs and $\mu_c$ to give twelve
predictions. The scenario III has powerful predictions with twelve
predictions. Scenario I and III are quite different methods. The
numerical results for the scenario I and III are similar and
corroborate each other.

The other schemes in Table~\ref{rsltcomp}  include the lattice
QCD~\cite{Can:2015exa}, the hyper central model~\cite{Patel:2007gx},
effective mass and screened charge scheme~\cite{Dhir:2009ax}, chiral
constituent quark model~\cite{Sharma:2010vv}, light-cone QCD sum
rules~\cite{Aliev:2008sk}, MIT bag
model~\cite{Bose:1980vy,Bernotas:2012nz}, Skyrmion
scheme~\cite{Oh:1991ws} and chiral quark-soliton
model~\cite{Yang:2018uoj}. Our results from all scenarios are less
than those from other schemes in general. Same tendency also
appeared in the magnetic moments of spin-$1\over 2$ charmed
baryon~~\cite{Wang:2018gpl}. In fact the lattice QCD results which
we used as input are also less than other schemes. In the lattice
QCD simulation, in order to extract the results with physical pion
mass, the rough linear or quadratic extrapolation was used in
Ref.~\cite{Can:2013tna}.

We have calculated the magnetic moments of spin-$3\over 2$ singly
heavy baryons analytically to $\mathcal{O}(p^3)$. The convergence of
the chiral expansion is good in our numerical results. For the lack
of experimental data, we have to adopt heavy quark symmetry and the
quark model to reduce and estimate our LECs. Our numerical results
can be improved with the new experimental results and the new
lattice QCD simulation results in the future. Meanwhile, our
analytical expressions can help the chiral extrapolation in lattice
QCD simulation. The LECs determined in this work can also be used to
study other physical properties, for instance, the electromagnetic
decay  of singly heavy baryon.

\section*{ACKNOWLEDGMENTS}

L. Meng is very grateful to H. S. Li, X. L. Chen and W. Z. Deng for
very helpful discussions. This project is supported by the National
Natural Science Foundation of China under Grants 11575008,
11621131001 and 973 program. This work is also supported by the
Fundamental Research Funds for the Central Universities of Lanzhou
University under Grants 223000--862637.

\begin{appendix}

\section{Integrals}\label{integrals}
We give some integrals with the conditions $v\cdot q=0$ and $q^2=0$.
All the results are given in the dimension $d=4$.
\begin{itemize}
\item Integrals with one meson propagator and one baryon propagator
\end{itemize}
\begin{equation}
i\int\frac{d^{d}l\,\lambda^{4-d}}{(2\pi)^{d}}\frac{l_{\alpha}l_{\beta}}{(l^{2}-m^{2}+i\epsilon)(\omega+v\cdot
l+i\epsilon)}=g_{\alpha\beta}J_{2}(\omega)+v_{\alpha}v_{\beta}J_{3}(\omega)
\end{equation}
\begin{equation}
J_{2}(\omega)=\begin{cases}
\frac{2\omega(\omega^{2}-m^{2})+\omega\left(3m^{2}-2\omega^{2}\right)\left(\ln\frac{m^{2}}{\lambda^{2}}+32\pi^{2}L(\omega)\right)-4\left(\omega^{2}-m^{2}\right)^{3/2}\left(\cosh^{-1}\left(\frac{\omega}{m}\right)-i\pi\right)}{16\pi^{2}(d-1)} & (\omega>m)\\
\frac{2\omega(\omega^{2}-m^{2})+\omega\left(3m^{2}-2\omega^{2}\right)\left(\ln\frac{m^{2}}{\lambda^{2}}+32\pi^{2}L(\omega)\right)+4\left(m^{2}-\omega^{2}\right)^{3/2}\cos^{-1}\left(-\frac{\omega}{m}\right)}{16\pi^{2}(d-1)} & (\omega^{2}<m^{2})\\
\frac{2\omega(\omega^{2}-m^{2})+\omega\left(3m^{2}-2\omega^{2}\right)\left(\ln\frac{m^{2}}{\lambda^{2}}+32\pi^{2}L(\omega)\right)+4\left(\omega^{2}-m^{2}\right)^{3/2}\cosh^{-1}\left(-\frac{\omega}{m}\right)}{16\pi^{2}(d-1)}
& (\omega<-m)
\end{cases}
\end{equation}

where $L(\lambda)$ is the infinite term:
\begin{equation}
L(\lambda)
=\frac{\lambda^{d-4}}{16\pi^{2}}\left[\frac{1}{d-4}-\frac{1}{2}\left(\text{ln}4\pi+1+\Gamma'(1)\right)\right]
\end{equation}

\begin{itemize}
\item Inegrals with two meson propagators and one baryon propagator
\end{itemize}
\begin{equation}
 i\int\frac{d^{d}l\,\lambda^{4-d}}{(2\pi)^{d}}\frac{l_{\alpha}l_{\beta}}{(l^{2}-m^{2}+i\epsilon)((l+q)^{2}-m^{2}+i\epsilon)(\omega+v\cdot l+i\epsilon)}=n_{1}^{\rmII}g_{\alpha\beta}+n_{2}^{\rmII}q_{\alpha}q_{\beta}+n_{3}^{\rmII}v_{\alpha}v_{\beta}+n_{4}^{\rmII}\left(v_{\alpha}q_{\beta}+q_{\alpha}v_{\beta}\right)
\end{equation}

\begin{equation}
 n_{1}^{\rmII}(\omega)=\begin{cases}
\frac{\omega\left(32\pi^{2}L(\lambda)+\ln\frac{m^{2}}{\lambda^{2}}\right)+2\sqrt{\omega^{2}-m^{2}}\left(\cosh^{-1}\left(\frac{\omega}{m}\right)-i\pi\right)}{8\pi^{2}(d-2)} & (\omega>m)\\
\frac{\omega\left(32\pi^{2}L(\lambda)+\ln\frac{m^{2}}{\lambda^{2}}\right)+2\sqrt{m^{2}-\omega^{2}}\cos^{-1}\left(-\frac{\omega}{m}\right)}{8\pi^{2}(d-2)} & (\omega^{2}<m^{2})\\
\frac{\omega\left(32\pi^{2}L(\lambda)+\ln\frac{m^{2}}{\lambda^{2}}\right)-2\sqrt{\omega^{2}-m^{2}}\cosh^{-1}\left(-\frac{\omega}{m}\right)}{8\pi^{2}(d-2)}
& (\omega<-m)
\end{cases}
\end{equation}

\begin{itemize}
\item Other integrals
\end{itemize}
The infinite terms are absorbed by the renormalization of the
coefficients and the $L(\lambda)$ term is omitted in the following
expression.

\begin{equation}
\frac{3-d}{d-1}n_{1}^{\rmII}(\omega,m)=\begin{cases}
-\frac{\omega\left(3\ln\frac{m^{2}}{\lambda^{2}}+1\right)+6\sqrt{\omega^{2}-m^{2}}\left[\cosh^{-1}\left(\frac{\omega}{m}\right)-i\pi\right]}{144\pi^{2}} & (\omega>m)\\
-\frac{\omega\left(3\ln\frac{m^{2}}{\lambda^{2}}+1\right)+6\sqrt{m^{2}-\omega^{2}}\cos^{-1}\left(-\frac{\omega}{m}\right)}{144\pi^{2}} & (\omega^{2}<m^{2})\\
-\frac{\omega\left(3\ln\frac{m^{2}}{\lambda^{2}}+1\right)-6\sqrt{\omega^{2}-m^{2}}\cosh^{-1}\left(-\frac{\omega}{m}\right)}{144\pi^{2}}
& (\omega<-m)
\end{cases}
\end{equation}

\begin{equation}
\frac{4}{d-1}n_{1}^{\rmII}(\omega,m)=\begin{cases}
\frac{\omega\left(3\ln\frac{m^{2}}{\lambda^{2}}-5\right)+6\sqrt{\omega^{2}-m^{2}}\left[\cosh^{-1}\left(\frac{\omega}{m}\right)-i\pi\right]}{36\pi^{2}} & (\omega>m)\\
\frac{\omega\left(3\ln\frac{m^{2}}{\lambda^{2}}-5\right)+6\sqrt{m^{2}-\omega^{2}}\cos^{-1}\left(-\frac{\omega}{m}\right)}{36\pi^{2}} & (\omega^{2}<m^{2})\\
\frac{\omega\left(3\ln\frac{m^{2}}{\lambda^{2}}-5\right)-6\sqrt{\omega^{2}-m^{2}}\cosh^{-1}\left(-\frac{\omega}{m}\right)}{36\pi^{2}}
& (\omega<-m)
\end{cases}
\end{equation}

\begin{equation}
J_{2}'(\omega)=\begin{cases}
\frac{\left(m^{2}-2\omega^{2}\right)\ln\left(\frac{m^{2}}{\lambda^{2}}\right)-4\omega\sqrt{\omega^{2}-m^{2}}\left[\cosh^{-1}\left(\frac{\omega}{m}\right)-i\pi\right]+2\omega^{2}}{16\pi^{2}} & (\omega>m)\\
\frac{\left(m^{2}-2\omega^{2}\right)\ln\left(\frac{m^{2}}{\lambda^{2}}\right)-4\omega\sqrt{m^{2}-\omega^{2}}\cos^{-1}\left(-\frac{\omega}{m}\right)+2\omega^{2}}{16\pi^{2}} & (\omega^{2}<m^{2})\\
\frac{\left(m^{2}-2\omega^{2}\right)\ln\left(\frac{m^{2}}{\lambda^{2}}\right)+4\omega\sqrt{\omega^{2}-m^{2}}\cosh^{-1}\left(-\frac{\omega}{m}\right)+2\omega^{2}}{16\pi^{2}}
& (\omega<-m)
\end{cases}
\end{equation}

\begin{equation}
\left(\frac{1-d}{4}+\frac{1}{d-1}\right)J_{2}'(\omega)=\begin{cases}
\frac{-15\left(m^{2}-2\omega^{2}\right)\ln\frac{m^{2}}{\lambda^{2}}+60\omega\sqrt{\omega^{2}-m^{2}}\left[\cosh^{-1}\left(\frac{\omega}{m}\right)-i\pi\right]-26m^{2}+22\omega^{2}}{576\pi^{2}} & (\omega>m)\\
\frac{-15\left(m^{2}-2\omega^{2}\right)\ln\frac{m^{2}}{\lambda^{2}}+60\omega\sqrt{m^{2}-\omega^{2}}\cos^{-1}\left(-\frac{\omega}{m}\right)-26m^{2}+22\omega^{2}}{576\pi^{2}} & (\omega^{2}<m^{2})\\
\frac{-15\left(m^{2}-2\omega^{2}\right)\ln\frac{m^{2}}{\lambda^{2}}-60\omega\sqrt{\omega^{2}-m^{2}}\cosh^{-1}\left(-\frac{\omega}{m}\right)-26m^{2}+22\omega^{2}}{576\pi^{2}}
& (\omega<-m)
\end{cases}
\end{equation}

\begin{equation}
\left(\frac{1-d}{2}+\frac{4}{d-1}-\frac{4}{(d-1)^{2}}\right)J_{2}'(\omega)=\begin{cases}
\frac{-33\left(m^{2}-2\omega^{2}\right)\ln\frac{m^{2}}{\lambda^{2}}+132\omega\sqrt{\omega^{2}-m^{2}}\left[\cosh^{-1}\left(\frac{\omega}{m}\right)-i\pi\right]-70m^{2}+74\omega^{2}}{864\pi^{2}} & (\omega>m)\\
\frac{-33\left(m^{2}-2\omega^{2}\right)\ln\frac{m^{2}}{\lambda^{2}}+132\omega\sqrt{m^{2}-\omega^{2}}\cos^{-1}\left(-\frac{\omega}{m}\right)-70m^{2}+74\omega^{2}}{864\pi^{2}} & (\omega^{2}<m^{2})\\
\frac{-33\left(m^{2}-2\omega^{2}\right)\ln\frac{m^{2}}{\lambda^{2}}-132\omega\sqrt{\omega^{2}-m^{2}}\cosh^{-1}\left(-\frac{\omega}{m}\right)-70m^{2}+74\omega^{2}}{864\pi^{2}}
& (\omega<-m)
\end{cases}
\end{equation}

\section{Renormaliztion}\label{renorm}
In the HBChPT, the divergences from the loops with fixed order
should be canceled out by renormalizing the LECs at this order. In
this section, we calculate the divergences of the loop diagrams and
give the renormalization of LECs explicitly. We take the mass
splittings as
\begin{equation}
\delta_1=0,\quad \delta_2=\delta_3\equiv \delta.
\end{equation}
We use the Gell-Mann-Okubo relation to express the $\eta$ mass,
$m_\eta^2=(4m_K^2-m_{\pi}^2)/3$. We adopt the leading order decay
constants for mesons $F_{\pi}=F_K=F_{\eta}\equiv F_{\phi}$ for
convenience.

The infinite parts of the $\mathcal{O}(p^3)$ loop diagrams (a) and
(b) are
\begin{equation}
\mathbb{L}^{(3)}_6\sim -{2\over
3F_\phi^2}g_b^2M_N\delta\mathbb{A}_0L(\lambda);\quad
\mathbb{L}^{(3)}_{6^*}\sim -{1\over
F_{\phi}^2}g_b^2M_N\delta\mathbb{A}_0L(\lambda).\label{infop3}
\end{equation}
where $\mathbb{L}$ denotes the infinite terms. The subscripts $6$
and $6^*$ denote the spin-$1\over 2$ and spin-$3\over 2$ sextets,
respectively. $\mathbb{A}_0=\{4,1,-2,1,-2,-2\}$ corresponds to the
loop coefficients of the sextet $\{\Sigma_{c}^{(*)++},
\Sigma_{c}^{(*)+}, \Sigma_{c}^{(*)0}, \Xi_{c}^{'(*)+},
\Xi_{c}^{'(*)0}, \Omega_{c}^{(*)0} \}$.  The infinite parts in
Eq.~(\ref{infop3}) for two sextets can be cancelled simultaneously
by one counter term,
\begin{equation}
{\cal L}_{{\cal H}\gamma}^{(3,ct)}=-i\frac{3g_{b}^{2}\delta
L(\lambda)}{2F_{\phi}^{2}}\langle\bar{{\cal
H}}^{\mu}\hat{F}_{\mu\nu}{\cal H}^{\nu}\rangle.
\end{equation}

 \begin{table}
\caption{The possible flavor structures of $\mathcal{O}(p^4)$
Lagrangian.
$(\chi_{+}f_{\mu\nu}^{+})_{ab}^{ij}\equiv(\chi_+)^{\{i}_{\{a}(f^+_{\mu\nu})^{j\}}_{b\}}$,
where the $\{...\}$ means that the flavor scripts are symmetrized.
}\label{flvstr}
\begin{tabular}{c|c|c|c|c|c|c|c}
\toprule[1pt]\toprule[1pt] Group representation  &
$1\times1\rightarrow1$ & $1\times8\rightarrow8$ &
$8\times1\rightarrow8$ & $8\times8\rightarrow1$ &
$8\times8\rightarrow8_{1}$ & $8\times8\rightarrow8_{2}$ &
$8\times8\rightarrow27$\tabularnewline\midrule[1pt] Flavor structure
& $\langle\tilde{\chi}_{+}\rangle\langle F_{\mu\nu}^{+}\rangle$ &
$\langle\tilde{\chi}_{+}\rangle\hat{F}_{\mu\nu}^{+}$ &
$\tilde{\chi}_{+}\langle F_{\mu\nu}^{+}\rangle$ &
$\langle\tilde{\chi}_{+}\hat{F}_{\mu\nu}^{+}\rangle$ &
$[\tilde{\chi}_{+},\hat{F}_{\mu\nu}^{+}]$ &
$\{\tilde{\chi}_{+},\hat{F}_{\mu\nu}^{+}\}$ &
$(\tilde{\chi}_{+}\hat{F}_{\mu\nu}^{+})_{\{a,b\}}^{\{i,j\}}$\tabularnewline
\hline LECs & $s_{5}/h_{5}/\kappa_{5}$ & $s_{6}/h_{6}/\kappa_{6}$ &
$s_{2}/h_{2}/\kappa_{2}$ & $s_{3}/h_{3}/\kappa_{3}$ & - &
$s_{1}/h_{1}/\kappa_{1}$ & $s_{4}/h_{4}/\kappa_{4}$\tabularnewline
\bottomrule[1pt]\bottomrule[1pt]
\end{tabular}
\end{table}

The divergences in the $\mathcal{O}(p^4)$ loop diagrams (e)-(l)
should be absorbed by the LECs in the $\mathcal{O}(p^4)$ chiral
Lagrangians. In Eq.~(\ref{4Bgamma}), we set $m_u=m_d=0$ and
$\chi_{+}=4B_0~\text{diag}(0,0,m_s)=4B_0m_s~\tilde{\chi}_+$. In this
section, we keep the $u/d$ quark mass. At the lead order, the
$\chi_+$ reads:
 \begin{equation}
 \chi_+=\text{diag}(m_{\pi}^{2},m_{\pi}^{2},2m_{K}^{2}-m_{\pi}^{2}),\quad \tilde{\chi}_+=\frac{1}{N}\text{diag}(m_{\pi}^{2},m_{\pi}^{2},2m_{K}^{2}-m_{\pi}^{2})
 \end{equation}
where $N=2m_{K}^{2}+m_{\pi}^{2}$. We have more novanishing
independent terms in the $\mathcal{O}(p^4)$ Lagrangians than those
in Eq.~(\ref{4Bgamma}). As illustrated in Table~\ref{flvstr}, there
are seven independent terms at this order.
$[\tilde{\chi}_{+},\hat{F}_{\mu\nu}^{+}]$ is vanishing at this order
since the leading terms of $\tilde{\chi}_{+}$ and
$\hat{F}_{\mu\nu}^{+}$ are both diagonal after the chiral expansion.
We can reconstruct the $\mathcal{O}(p^4)$ Lagrangians for
spin-$1\over 2$ and spin-$3\over 2$ sextets, respectively,
 \begin{eqnarray}
 {\cal L}_{B\gamma}^{(4)} &=&\frac{s_{1}}{8M_{N}}\langle\bar{B}_{6}\sigma^{\mu\nu}\{\tilde{\chi}_{+},\hat{F}_{\mu\nu}^{+}\}B_{6}\rangle+\frac{s_{2}}{8M_{N}}\langle\bar{B}_{6}\sigma^{\mu\nu}\tilde{\chi}_{+}B_{6}\rangle\langle F_{\mu\nu}^{+}\rangle+\frac{s_{3}}{8M_{N}}\langle\bar{B}_{6}\sigma^{\mu\nu}B_{6}\rangle\langle\tilde{\chi}_{+}\hat{F}_{\mu\nu}^{+}\rangle\nonumber \\
    &&+\frac{s_{4}}{2M_{N}}\langle\bar{B}_{6}\sigma^{\mu\nu}\hat{F}_{\mu\nu}^{+}B_{6}\tilde{\chi}_{+}^{T}\rangle+\frac{s_{5}}{8M_{N}}\langle\bar{B}_{6}\sigma^{\mu\nu}B_{6}\rangle\langle\tilde{\chi}_{+}\rangle\langle F_{\mu\nu}^{+}\rangle+\frac{s_{6}}{8M_{N}}\langle\bar{B}_{6}\sigma^{\mu\nu}\langle\tilde{\chi}_{+}\rangle\hat{F}_{\mu\nu}^{+}B_{6}\rangle,\\
{\cal L}_{B^{*}\gamma}^{(4)}&=&\frac{ih_{1}}{4M_{N}}\langle\bar{B}_{6}^{*\mu}\{\tilde{\chi}_{+},\hat{F}_{\mu\nu}^{+}\}B_{6}^{*\nu}\rangle+\frac{ih_{2}}{4M_{N}}\langle\bar{B}_{6}^{*\mu}\tilde{\chi}_{+}B_{6}^{*\nu}\rangle\langle F_{\mu\nu}^{+}\rangle+\frac{ih_{3}}{4M_{N}}\langle\bar{B}_{6}^{*\mu}B_{6}^{*\nu}\rangle\langle\tilde{\chi}_{+}\hat{F}_{\mu\nu}^{+}\rangle \nonumber\\
&
&+\frac{ih_{4}}{M_{N}}\langle\bar{B}_{6}^{*\mu}\hat{F}_{\mu\nu}^{+}B_{6}^{*\nu}\tilde{\chi}_{+}^{T}\rangle+\frac{ih_{5}}{4M_{N}}\langle\bar{B}_{6}^{*\mu}B_{6}^{*\nu}\rangle\langle\tilde{\chi}_{+}\rangle\langle
F_{\mu\nu}^{+}\rangle+\frac{ih_{6}}{4M_{N}}\langle\bar{B}_{6}^{*\mu}\hat{F}_{\mu\nu}^{+}B_{6}^{*\nu}\rangle\langle\tilde{\chi}_{+}\rangle.
 \end{eqnarray}
  The $h_2(s_2)$ and $h_5(s_5)$ terms involve the $\langle F_{\mu\nu}^{+}\rangle$. The two terms represent the heavy quark's contribution. The other terms represent the light quarks' contribution. The magnetic moments of the spin-$1\over 2$ and spin-$3\over 2$ sextets are
  \begin{eqnarray}
  \mu_6=\sum_{i=1}^{6}s_i \mathbb{A}_i, \quad \mu_{6^*}=-\sum_{i=1}^{6}h_i\mathbb{A}_i,\label{mmg}
\end{eqnarray}
where $\mathbb{A}_i$ is the coefficients of $h_i$ or $s_i$ in
Table~\ref{Ai}. The divergences in the loop diagrams of spin-$1\over
2$ and spin-$3\over 2$ sextets can be absorbed by $s_i$ and $h_i$,
respectively.

\begin{table}
\caption{Coefficients of the magnetic moments that arise from the
$\mathcal{O}(p^4)$ Lagrangians. The $\mathbb{A}_i$ is also used to
express the infinite parts of the $\mathcal{O}(p^4)$ loop
diagrams.}\label{Ai}
\begin{tabular}{c|c|c|c|c|c|c}
\toprule[1pt]\toprule[1pt]
 & $\Sigma_{c}^{(*)++}$ & $\Sigma_{c}^{(*)+}$ & $\Sigma_{c}^{(*)0}$ & $\Xi_{c}^{'(*)+}$ & $\Xi_{c}^{'(*)0}$ & $\Omega_{c}^{(*)0}$\tabularnewline
\midrule[1pt] $\mathbb{A}_{1}$ & $\frac{4}{3N}m_{\pi}^{2}$ &
$\frac{1}{3N}m_{\pi}^{2}$ & $-\frac{2}{3N}m_{\pi}^{2}$ &
$(-\frac{2}{3N}m_{K}^{2}+\frac{1}{N}m_{\pi}^{2})$ &
$-\frac{2}{3N}m_{K}^{2}$ &
$(-\frac{4}{3N}m_{K}^{2}+\frac{2}{3N}m_{\pi}^{2})$\tabularnewline
$\mathbb{A}_{2}$ & $\frac{m_{\pi}^{2}}{N}$ & $\frac{m_{\pi}^{2}}{N}$
& $\frac{m_{\pi}^{2}}{N}$ & $\frac{m_{K}^{2}}{N}$ &
$\frac{m_{K}^{2}}{N}$ &
$\frac{2m_{K}^{2}-m_{\pi}^{2}}{N}$\tabularnewline $\mathbb{A}_{3}$ &
$\frac{2}{3N}(m_{K}^{2}-m_{\pi}^{2})$ &
$\frac{2}{3N}(m_{K}^{2}-m_{\pi}^{2})$ &
$\frac{2}{3N}(m_{K}^{2}-m_{\pi}^{2})$ &
$\frac{2}{3N}(m_{K}^{2}-m_{\pi}^{2})$ &
$\frac{2}{3N}(m_{K}^{2}-m_{\pi}^{2})$ &
$\frac{2}{3N}(m_{K}^{2}-m_{\pi}^{2})$\tabularnewline
$\mathbb{A}_{4}$ & $\frac{8}{3N}m_{\pi}^{2}$ &
$\frac{2}{3N}m_{\pi}^{2}$ & $-\frac{4}{3N}m_{K}^{2}$ &
$(\frac{8}{3N}m_{K}^{2}-\frac{2}{N}m_{\pi})$ &
$-\frac{4}{3N}m_{K}^{2}$ &
$(-\frac{8}{3}m_{K}^{2}+\frac{4}{3N}m_{\pi}^{2})$\tabularnewline
$\mathbb{A}_{5}$ & 1 & 1 & 1 & 1 & 1 & 1\tabularnewline
$\mathbb{A}_{6}$ & $\frac{2}{3}$ & $\frac{1}{6}$ & $-\frac{1}{3}$ &
$\frac{1}{6}$ & $-\frac{1}{3}$ & $-\frac{1}{3}$\tabularnewline
\bottomrule[1pt]\bottomrule[1pt]
\end{tabular}
\end{table}

The LECs $h_i$ and $s_i$ can be related to each other using the
heavy quark expansion. In the heavy quark limit, the $\mathcal
O(p^4)$ Lagrangians read

 \begin{eqnarray}
 {\cal L}_{\mathcal{B}\gamma}^{(4)} &=&\frac{i\kappa_{1}}{4M_{N}}\langle\bar{{\cal H}^{\mu}}\{\tilde{\chi}_{+},\hat{F}_{\mu\nu}^{+}\}{\cal H^{\nu}}\rangle+\frac{i\kappa_{3}}{4M_{N}}\langle\bar{{\cal H}^{\mu}}{\cal H^{\nu}}\rangle\langle\tilde{\chi}_{+}\hat{F}_{\mu\nu}^{+}\rangle \nonumber\\
    & &+\frac{i\kappa_{4}}{M_{N}}\langle\bar{{\cal H}^{\mu}}\hat{F}_{\mu\nu}^{+}{\cal H^{\nu}}\tilde{\chi}_{+}^{T}\rangle+\frac{i\kappa_{6}}{4M_{N}}\langle\bar{{\cal H}^{\mu}}\hat{F}_{\mu\nu}^{+}{\cal H^{\nu}}\rangle\langle\tilde{\chi}_{+}\rangle,
 \end{eqnarray}
where ${\cal H}^{\mu}$ is the superfield defined in
Eq.~(\ref{superfield}). The heavy quark symmetry breaking
Lagrangians at the order of ${1\over m_c}$ read
   \begin{eqnarray}
 {\cal {\cal L}}_{HQ}^{(4)}=\frac{\kappa_{2}}{8M_{N}}\langle\bar{{\cal H}}^{\rho}\sigma_{\mu\nu}\tilde{\chi}_{+}{\cal H}_{\rho}\rangle\langle F^{+\mu\nu}\rangle+\frac{\kappa_{5}}{8M_{N}}\langle\bar{{\cal H}}^{\rho}\sigma_{\mu\nu}\tilde{\chi}_{+}{\cal H}_{\rho}\rangle\langle\tilde{\chi}_{+}\rangle\langle F^{+\mu\nu}\rangle,
\end{eqnarray}
The ${\cal L}_{\mathcal{B}\gamma}^{(4)}$ and $ {\cal {\cal
L}}_{HQ}^{(4)}$ describe the dynamics of the light quark and the
heavy quark sectors, respectively. The $s_i$ and $h_i$ are related
to the $\kappa_i$ as
\begin{eqnarray}
s_{i}=\begin{cases}
-\frac{2}{3}\kappa_{i} & i=1,3,4,6\\
\frac{1}{3}\kappa_{i} & i=2,5
\end{cases},\quad h_{i}=\begin{cases}
\kappa_{i} & i=1,3,4,6\\
\kappa_{i} & i=2,5\label{relation}
\end{cases},\Rightarrow
s_{i}=\begin{cases}
-\frac{2}{3}h_{i} & i=1,3,4,6\\
\frac{1}{3}h_{i} & i=2,5
\end{cases}
\end{eqnarray}
Since the $\mathcal{O}(p^4)$ LECs of the spin-$1\over 2$ and
spin-$3\over 2$ are related to each other as shown in
Eq.~(\ref{relation}), we expect the divergences of the
$\mathcal{O}(p^4)$ loop diagrams have the same relations.

Here we give the infinite parts of loop diagrams explicitly. For the
spin-$1\over 2$ sextet, the divergences of the loops at $\mathcal
O^(p^4)$ are
\begin{eqnarray}
\mathbb{L}_{6}^{(4)}\sim
\sum_{i=1,3,4,6}a_i\mathbb{A}_iL(\lambda)+\sum_{i=2,5}a_i\mathbb{A}_iL(\lambda)\label{dive6}
\end{eqnarray}
with
\begin{eqnarray}
a_{1}   &=&\frac{N}{12F_{\phi}^{2}}\left[g_{c}\left(8g_{a}^{2}+3g_{b}^{2}+6\right)-\frac{16}{3}g_{a}g_{b}g_{e}+3g_{f}\right],\nonumber\\
a_{2}   &=&-\frac{N}{6F_{\phi}^{2}}g_{b}^{2}\left(2d_{3}+g_{H}\right),\nonumber\\
a_{3}   &=&\frac{N}{12F_{\phi}^{2}}\left[-8g_{a}g_{b}g_{e}+\left(3g_{a}^{2}+4\right)g_{c}+2g_{f}\right],\nonumber\\
a_{4}   &=&\frac{N}{48F_{\phi}^{2}}\left(-\frac{40}{3}g_{a}g_{b}g_{e}+7g_{a}^{2}g_{c}+6g_{b}^{2}g_{c}\right),\nonumber\\
a_{5}   &=&\frac{8\delta^{2}-N}{6F_{\phi}^{2}}g_{b}^{2}\left(2d_{3}+g_{H}\right),\nonumber\\
a_{6}
&=&\frac{N}{36F_{\phi}^{2}}\left[g_{c}\left(35g_{a}^{2}+36g_{b}^{2}+12\right)-24g_{a}g_{b}g_{e}+6g_{f}\right]+\frac{8}{3F_{\phi}^{2}}g_{a}g_{b}g_{e}\delta^{2},
\end{eqnarray}
where the $a_2$ and $a_5$ are related to the $d_3$ and $g_H$, which
are coupling constants of the heavy quark sector. The other terms
only involve the coupling constants of the light quark sector.

For the spin-$3\over 2$, the infinite parts of $\mathcal{O}(p^4)$
loop diagrams read
\begin{eqnarray}
\mathbb{L}_{6^*}^{(4)}\sim \sum_{i=1,3,4,6}{3\over
2}a_i\mathbb{A}_iL(\lambda)-\sum_{i=2,5}3a_i\mathbb{A}_iL(\lambda).\label{dive6s}
\end{eqnarray}
We  find that
\begin{eqnarray}
\mathbb{L}_{6,l}^{(4)}={2\over 3}\mathbb{L}_{6^*,l}^{(4)},\quad
\mathbb{L}_{6,h}^{(4)}=-{1\over
3}\mathbb{L}_{6^*,h}^{(4)},\label{66srelated}
\end{eqnarray}
where the subscripts $l$ and $h$ represent the contributions from
the light quarks and heavy quark, respectively. The divergences of
the chiral loops respect the relations between $s_i$ and $h_i$ in
the heavy quark limit (up to a $-1$ factor arising from
Eq.~(\ref{mmg})). Even though the number of the LECs are reduced
using the heavy quark symmetry, the divergences of the two sextets
can be renormalized simultaneously. We give the renormalization of
$\kappa_i$ explicitly,
\begin{eqnarray}
\kappa_{i}(\lambda)=\begin{cases}
\kappa_{i}^{(r)}+\frac{3}{2}a_{i}L(\lambda) & i=1,3,4,6\\
\kappa_{i}^{(r)}-3a_{i}L(\lambda) & i=2,5
\end{cases}.
\end{eqnarray}

\end{appendix}

\vfil \thispagestyle{empty}
\newpage
\bibliography{ref}

\end{document}